\documentclass[12pt,preprint]{aastex}
\usepackage{epsf}
\newcommand{\simlt}{\lower.5ex\hbox{$\; \buildrel < \over \sim \;$}}
\newcommand{\bm}[1]{{\mbox{\boldmath${#1}$}}}
\newcommand{\iint}{\int\!\!\!\int}
\slugcomment{}
\shortauthors{Ohta, Taruya \& Suto}
\shorttitle{Rossiter -- McLaughlin effect for extrasolar planetary systems}
\begin{document}
%
%
\title{The Rossiter -- McLaughlin effect and analytic radial velocity
 curves for  transiting extrasolar planetary systems}
\author{Yasuhiro Ohta, Atsushi Taruya\altaffilmark{1}
and Yasushi Suto\altaffilmark{1}}
\affil{Department of Physics, The University of Tokyo, 
 Tokyo 113-0033, Japan}
\email{ohta@utap.phys.s.u-tokyo.ac.jp, ataruya@utap.phys.s.u-tokyo.ac.jp,
suto@phys.s.u-tokyo.ac.jp}
\altaffiltext{1}{Also at Research Center for the Early Universe(RESCEU), School
of Science, The University of Tokyo, Tokyo 113-0033, Japan.}
%
%
\begin{abstract}
 A transiting extrasolar planet sequentially blocks off the light coming
 from the different parts of the disk of the host star in a time
 dependent manner. Because of the spin of the star, this produces an
 asymmetric distortion in the line profiles of the stellar spectrum,
 leading to an apparent anomaly in the radial velocity curves, known as
 the Rossiter -- McLaughlin effect.  Here, we derive approximate but
 accurate analytic formulae for the anomaly in the radial velocity curves,
 taking into account the stellar limb darkening. The formulae are
 particularly useful in extracting information on the projected angle
 between the planetary orbit axis and the stellar spin axis, $\lambda$,
 and the projected stellar spin velocity, $V\sin I_s$.  We create mock
 samples for the radial curves for the transiting extrasolar system
 HD 209458 and demonstrate that constraints on the spin parameters
 $(V\sin I_s,~\lambda)$ can be significantly improved by combining our
 analytic template formulae and the precision velocity curves from
 high-resolution spectroscopic observations with 8--10 m class
 telescopes. Thus, future observational exploration of transiting systems
 using the Rossiter -- McLaughlin effect will be one of the most important
 probes for a better understanding of the origin of extrasolar planetary
 systems, especially the origin of their angular momentum.
\end{abstract}
\keywords{planets and satellites: individual (HD 209458b),  
techniques: spectroscopic}
%
%
\section{Introduction}

With more than 130 extrasolar planets discovered so far, major
scientific purposes in this field are rapidly moving from mere detection
to characterization of the planetary systems, i.e., statistics of
planetary masses, orbital periods, eccentricities, stellar metallicities,
and so on.  Most of these systems have been discovered through the periodic
change of radial velocities of the central stars. One of them, HD 209458b,
was first discovered spectroscopically and was soon later found to exhibit a
transit signature in front of the stellar disk for a duration of $\sim
2$ hr in its orbital period of 3.5 days \citep{henry00,Charbonneau00}.
More recently, a few additional extrasolar planets have been discovered
photometrically from a transiting signature in their
light curves and later confirmed spectroscopically
\citep[e.g.,][]{Udalski02c,Bouchy,Alonso}.

Indeed, such transiting planets provide important information for the
extrasolar planetary systems that is otherwise unavailable: planetary size,
atmospheric composition, and the degree of (mis)alignment of the
planetary orbit axis and the stellar spin axis.  Among others,
\citet{Queloz00} showed that the planetary orbit and the stellar
rotation of the HD 209458 system share the same direction; the planet
sequentially blocks off the light from the approaching and then from the
receding parts of the stellar surface. This produces a distortion in the
line profiles of the stellar spectrum during the transit in a
time-dependent manner, leading to an anomaly of the radial velocity
curves, previously known as the Rossiter -- McLaughlin (RM)
effect in eclipsing binary stars
\citep{Rossiter24,McLaughlin24,Kopal99}. 
The signature of the RM effect was first reported 
by \citet{Schlesinger1909} and the effect was later 
isolated and measured by \citet{Rossiter24} and \citet{McLaughlin24}.
The first attempt at a theoretical study of 
the RM effect was made by \citet{Petrie38} and the analysis was 
subsequently  extended by \citet{Kopal42,Kopal45} 
to incorporate the effects of limb and gravity
darkening and rotational and tidal distortion. 
In the context of extrasolar planetary systems, 
\citet{Queloz00} numerically computed the expected amplitude of the
radial velocity anomaly caused by the RM effect. They first put constraints 
on the stellar spin angular velocity and
its direction angle with respect to the planetary orbit by comparing
the observed velocity anomalies. Incidentally,
\citet{Snellen04} proposed an interesting suggestion to use the
RM effect as an efficient diagnostic of the atmosphere of 
a transiting planet.

Obviously, this methodology provides unique and fundamental clues to
understanding the formation process of extrasolar planetary systems.
Planets are supposed to form in the proto planetary disk surrounding the
proto star \citep[e.g.,][]{Pollack96}. Thus, the stellar spin and the
planetary orbital axes are expected to be aligned. In turn, any
constraints on their (mis)alignment degree are useful clues to the
origin of the angular momentum of planets and its subsequent evolution
during possible migration of the planets into the close-in orbits
\citep[e.g.,][]{Lin96}. In order to improve the reliability and precision
of such results, analytic templates are of great value.  In this
paper, we show analytic formulae for radial velocities of transiting
extrasolar planets.  Those formulae can be used as standard templates in
constraining a set of parameters for numerous transiting planets that
will be detected in the near future with {\it Corot} and {\it Kepler}.

The rest of the paper is organized as follows. Section
\ref{sec:radvel_non-transit} introduces a variety of parameters that
characterize the dynamics of a planetary system and summarizes the
radial velocity curve, neglecting the transit effect (or, equivalently, for
a non rotating star). Section \ref{sec:radvel_transit} shows
a general theoretical framework of the RM effect for a star during the
planetary transit. Sections \ref{sec:without_darkening} and
\ref{sec:limb_darkening} derive analytic expressions for radial velocity
curves; in \S \ref{sec:without_darkening}, we consider an idealistic
case of the stellar intensity model without limb darkening and derive
the exact analytic expressions. Notation used in the derivation of are
summarized in Table 1. On this basis, \S
\ref{sec:limb_darkening} presents approximate formulae taking into
account the stellar limb-darkening effect.  We apply these analytic
templates to a transiting system, HD 209458, in \S \ref{sec:HD209458} and
examine the sensitivity to the parameters of the system.  Finally, \S
\ref{sec:conclusion} is devoted to the main conclusions and discussion.

\section{Radial velocity profile for a star with a nontransiting planet}
\label{sec:radvel_non-transit}

\begin{figure}[htb]
 \epsscale{0.5}
 \plotone{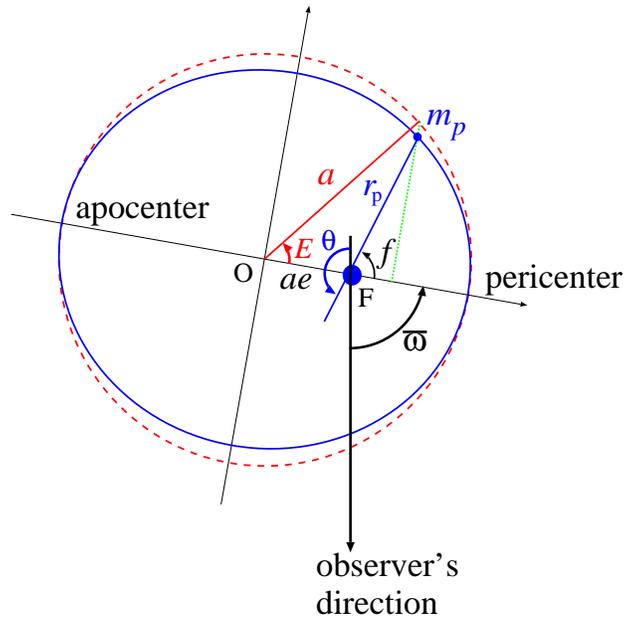}
\caption{Schematic top view of the planetary orbit. The star is
 located at the focus point F (see Table \ref{tab:notation} for
 meaning of symbols).
\label{fig:orbit-topview}}
\end{figure}

A close-in extrasolar planetary system may have multiple outer planets, but
we focus here on a system that is well approximated by a two-body problem,
i.e., one that consists of a central star (mass $m_s$) and a planet (mass
$m_p$). Figure \ref{fig:orbit-topview} shows the schematic configuration
of the top view of the planetary orbit.  The radial velocity curve of
the star in the Kepler orbit can be described as follows
\citep[e.g.,][]{SSD}.

First, note that in the strictly two-body problem, the orbit of the
planet with respect to the star is simply written as
\begin{eqnarray}
\label{eq:rp}
 r_p = \frac{a(1-e^2)}{1+e\cos f} ,
\end{eqnarray}
where $a$ is the semimajor axis, $e$ is the eccentricity, and $f$ is the
true anomaly (angular coordinate measured from the pericenter
direction).  The true anomaly $f$ is written in terms of the eccentric
anomaly $E$, defined through the circumscribed circle that is concentric
with the orbital ellipse as
\begin{eqnarray}
\label{eq:true_anomaly}	
 \cos f = \frac{\cos E -e}{1-e\cos E} .
\end{eqnarray}
If one introduces the mean motion $n$ from the orbital period $P_{\rm orb}$ of
the system as
\begin{eqnarray}
\label{eq:mean_motion}
n \equiv \frac{2\pi}{P_{\rm orb}}, 
\end{eqnarray}
then $E$ is related to the mean anomaly $M$ as (Kepler's equation)
\begin{eqnarray}
\label{eq:mean_anomaly}
M= E -e \sin E ,
\end{eqnarray}
where $M \equiv n(t-\tau)$, where $\tau$ is the time of pericenter
passage. 

Using the parameters defined above, the radial velocity of the star
along the line of sight of the observer (see Fig.
\ref{fig:orbit-topview}) is written as \citep[e.g.,][]{SSD}
\begin{eqnarray}
\label{eq:radvel_star}
 v_{\rm rad,s}=-\frac{m_p}{m_s+m_p}\frac{n a\sin i}{\sqrt{1-e^2}}
 \left[\sin(f+\varpi)+e\sin\varpi \right] ,
\end{eqnarray}
where $i$ denotes the inclination angle between the direction normal to
the orbital plane and the observer's line of sight and we define
$-\varpi$ as the longitude of the line of sight with respect to the
pericenter (Fig.\ref{fig:orbit-topview}). While $f$ is not directly
written as a function of the observer's time $t$, it is useful to
rewrite equation (\ref{eq:radvel_star}) explicitly in terms of $M =
n(t-\tau)$ even in an approximate manner.  For this purpose, one can use
the following expansions with respect to the eccentricity $e$
\citep[e.g.,][]{SSD}:
\begin{eqnarray}
\sin f &=& 2\sqrt{1-e^2} \sum_{k=1}^\infty \frac{1}{k} \frac{d}{de}
J_k(ke) \sin kM \cr
&=& \sin M + e\sin 2M 
+ e^2 \left(\frac{9}{8} \sin 3M - \frac{7}{8}\sin M \right) + O(e^3), \\
\cos f &=& -e + \frac{2(1-e^2)}{e} \sum_{k=1}^\infty J_k(ke) \cos kM \cr
&=& \cos M + e(\cos 2M -1) + \frac{9e^2}{8}(\cos 3M -\cos M)  + O(e^3).
\end{eqnarray}
Then equation (\ref{eq:radvel_star}) up to $O(e)$ should read
\begin{eqnarray}
\label{eq:radvel_star_e}
 v_{\rm rad,s}  \approx -\frac{m_p n a\sin i}{m_s+m_p}
\left[\sin(M+\varpi) + e \sin (2M+\varpi)  \right] .
\end{eqnarray}

\section{Radial velocity profile for a star with a transiting planet}
\label{sec:radvel_transit}

\begin{figure}[htb]
\epsscale{0.5}
 \plotone{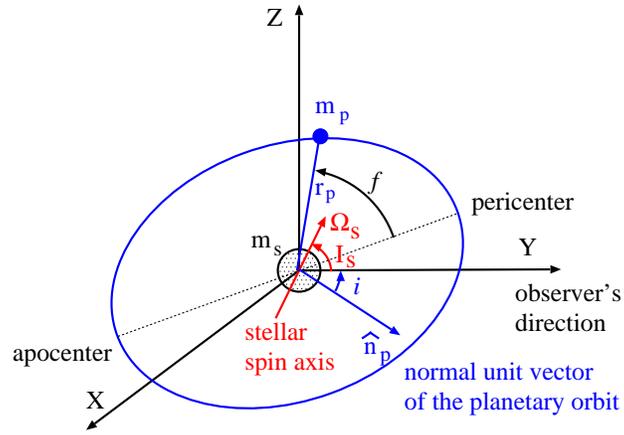}
 \caption{Schematic configuration of the stellar spin axis, the
 planetary orbital plane, and the observer's line of sight (see Table
 \ref{tab:notation} for meaning of symbols).  \label{fig:spin-los}}
\end{figure}

\begin{figure}[htb]
\epsscale{0.5}
 \plotone{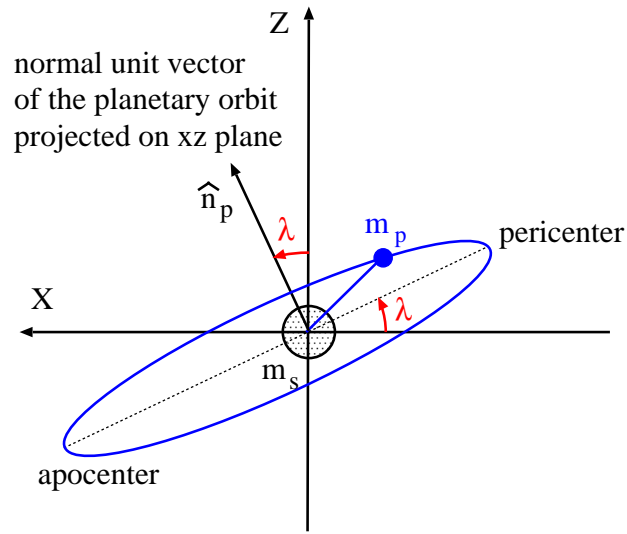}
\caption{Projected view of the planetary orbital plane 
from the line of sight ($Y$-axis in this case).
\label{fig:orbit-xz}}
\end{figure}

An occultation of a part of the rotating stellar surface during transit
of the planet causes a time-dependent asymmetric feature in stellar
emission/absorption line profiles. If the line profile is not well
resolved, the asymmetry results in an apparent shift of the central line
position, which contributes additionally to the overall ``observed''
stellar radial velocity. In order to describe the effect quantitatively,
we set the coordinate system centered at the star so that its $y$-axis
is directed toward the observer (Fig.\ref{fig:spin-los}). The $z$-axis
is chosen so that the stellar rotation axis lies on the $y$-$z$ plane.  We
also define the angle $\lambda$ between the $z$-axis and the normal vector
$\hat{\bm{n}}_{p}$ of the planetary orbit plane projected on the $x$-$z$ plane
(Fig.\ref{fig:orbit-xz}), i.e.,
\begin{eqnarray}
\label{eq:hat-np}
\hat{\bm{n}}_{p} = \left( 
\begin{array}{c}
       \sin\lambda\sin i \\
       \cos i \\
       \cos\lambda\sin i \\
\end{array}
\right) .
\end{eqnarray}
Then the position of the planet is given by
\begin{eqnarray}
\label{eq:planet_position}
 \bm{X}&=&R_y(\lambda)R_x\left(i-\frac{\pi}{2}\right)R_z
\left(\varpi+\frac{\pi}{2}\right)
\left(\begin{array}{c} r_p\cos f\\ r_p\sin f\\ 0 \end{array}\right)
\cr
 &=&r_p\left(
\begin{array}{c}
 -\cos\lambda\sin(f+\varpi)-\sin\lambda\cos i\cos(f+\varpi)\\
 \sin i\cos(f+\varpi)\\
 \sin\lambda\sin(f+\varpi)-\cos\lambda\cos i\cos(f+\varpi)
\end{array}\right) ,
\end{eqnarray}
where $R_k(\theta)$ denotes the rotation matrix of an angle $\theta$ 
around the $k$-axis.

In our configuration, the angular velocity of the star is given as 
\begin{eqnarray}
{\bf \Omega}_s = (0, \Omega_s \cos I_s, \Omega_s \sin I_s). 
\end{eqnarray}
Then the velocity of a point $\bm{R}=(x,y,z)$ on the stellar surface is
\begin{eqnarray}
 \bm{v} = {\bf \Omega}_s\times\bm{R}=\Omega_s\left(
  \begin{array}{c}
   z\cos I_s - y\sin I_s \\
   x\sin I_s \\
   -x\cos I_s
  \end{array}\right) .
\label{eq:angular_veloc}
\end{eqnarray}
Thus, the radiation at frequency $\nu$ from that point suffers from the
Doppler shift due to the stellar rotation by an amount
\begin{eqnarray}
\label{eq:doppler-shift}
\frac{\Delta\nu}{\nu}= \frac{\Omega_s x \sin I_s}{c}
\end{eqnarray}
with respect to the observer located along the $y$-axis in the present
case.

Consider a specific (emission or absorption) line whose intensity at a
point $(x,z)$ on the projected stellar surface is given by
$I_\nu(x,z)=I(x,z)H(\nu)$, where $H(\nu)$ represents the line profile.
The observed flux is computed by integrating the Doppler-shifted
intensity at each point over the entire (projected) surface of the star:
\begin{eqnarray}
\label{eq:doppler-flux}
 F_\nu = \int \left(1+ \frac{\Delta\nu}{\nu}\right)^3
I(x,z)H(\nu-\Delta\nu) \frac{dx dz}{D^2} ,
\end{eqnarray}
where $D$ is the distance between the star and the observer.  The factor
$(1+\Delta\nu/\nu)^3$ appears because of the Lorentz invariance of the
quantity $I_{\nu}/\nu^3$. While our analysis is applicable to both
emission and absorption lines, we consider an emission line centered at
$\nu=\nu_0$ in the following, just for definiteness.  Then the line
profile function satisfies
\begin{eqnarray}
 \int H(\nu)d\nu&=&1 , 
\label{eq:line_profile1}\\ 
 \int \nu H(\nu)d\nu&=&\nu_0. 
\label{eq:line_profile2}
\end{eqnarray}
Since $H(\nu)$ is supposed to be sharply peaked only around $\nu_0$, we
have approximately
\begin{equation}
 \int f(\nu) H(\nu)d\nu\approx f(\nu_0) ,
\label{eq:line_profile3}
\end{equation}
for an arbitrary smooth function $f(\nu)$.

If the resolution of the observational spectrograph were sufficiently
high, the line profiles of the star and the planetary shadow would be
separated, or at least the asymmetric feature might be detected for
transiting systems \citep[e.g.,][]{Charbonneau98,Charbonneau99}. In
reality, however, such a high spectral resolution is quite demanding,
and here we assume a somewhat lower resolution. Thus, we simply compute
the resulting time-dependent shift of the line-profile-weighted mean
position $\bar\nu$ due to an asymmetric occultation of the stellar
surface during the passage of the transiting planet. Using
expression (\ref{eq:doppler-flux}) and the properties of the line profile
function (eqs. [\ref{eq:line_profile1}] to [\ref{eq:line_profile3}]), we
obtain
\begin{eqnarray}
 \bar{\nu}
&\equiv& \frac{\displaystyle \int\nu F_\nu d\nu}{\displaystyle 
\int F_\nu d\nu} \cr 
&=& \nu_0\left\{ 1 + \frac{\displaystyle 
\iint\,\left(1+\frac{\Delta\nu}{\nu_0}\right)^3 
\frac{\Delta\nu}{\nu_0}\,\,I(x,z)dxdz}
{\displaystyle 
\iint \left(1+\frac{\Delta\nu}{\nu_0}\right)^3\,\,I(x,z)dxdz}\right\}.
\end{eqnarray}
Since the amplitude of the Doppler shift (eq.[\ref{eq:doppler-shift}]) 
is small, one can safely expand $\bar \nu$ up to the leading order of 
$\Delta\nu/\nu$ as 
\begin{eqnarray}
 \bar{\nu} &=& \nu_0\left\{ 1+\frac{\displaystyle 
\iint \frac{\Delta\nu}{\nu_0}\,\,I(x,z)dxdz}
{\displaystyle \iint I(x,z)dxdz}+{\cal O}
\left(\frac{\Delta\nu^2}{\nu_0^2}\right)\right\} \cr
 &\approx& \nu_0\left\{1+\frac{\Omega_s\sin I_s}{c}\,\,
\frac{\displaystyle \iint x\,I(x,z)dxdz}
{\displaystyle  \iint I(x,z)dxdz}\right\} .
\end{eqnarray}
Therefore, the ``apparent'' radial stellar velocity anomaly due to the RM
effect is expressed as
\begin{eqnarray}
\label{eq:delta_vs}
\Delta v_{s} =
-\Omega_s\sin I_s
\frac{\displaystyle \iint x\,I(x,z)dxdz}
{\displaystyle \iint I(x,z)dxdz} .
\end{eqnarray}

\begin{figure}[htb]
 \epsscale{1.0} \plottwo{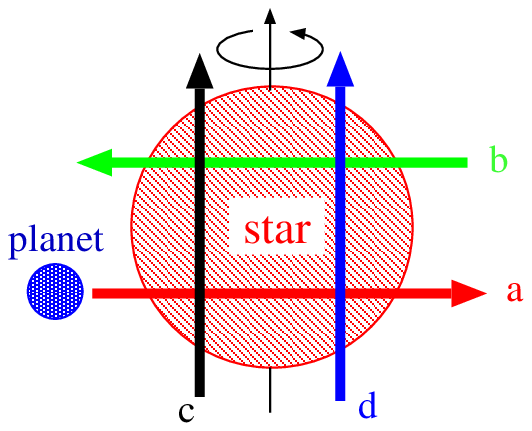}{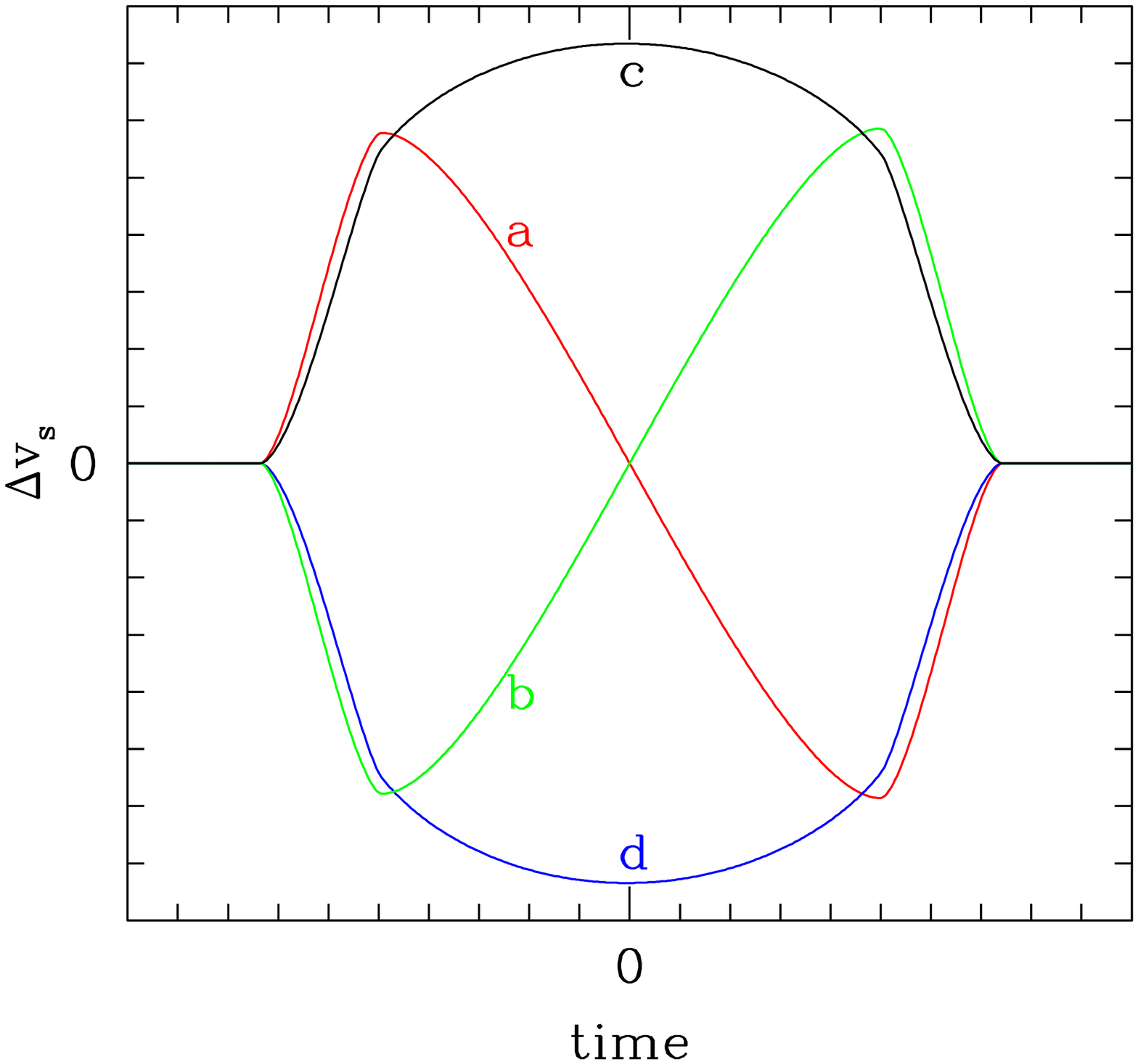} \caption{Schematic
 illustration of the velocity curve anomaly due to the RM effect. The
 four different paths of a planet, a-d, in the left panel
 correspond to the velocity curves in the right panel.
 \label{fig:radvel360}}
\end{figure}

Equation (\ref{eq:delta_vs}) is the basic relation between $I(x,z)$ and
$\Delta v_s$ in our subsequent analysis.  Figure \ref{fig:radvel360}
shows a schematic illustration of the RM effect. Depending on the
inclination and the orbital rotation direction relative to the stellar
spin axis, the velocity curve anomaly due to the RM effect exhibits
rather different behavior.

The remaining task is to evaluate the integrals adopting a certain model
of a stellar surface intensity $I(x,z)$.  Note that previous literature
in the analytic study of radial velocity curves focused on expressing
the integrals in the radial velocity shift (\ref{eq:delta_vs}) in terms
of Kopal's associated $\alpha$-functions under some model assumptions of
the stellar intensity \citep[e.g.,][]{Hosokawa53,Kopal99}.  While such
detailed and exact approaches are required for stellar eclipsing
binaries, the evaluation of the $\alpha$-function is a demanding
numerical task. Furthermore, for those systems a variety of effects
become important, including limb darkening, distortion of stars due to
their rotation and tidal interaction, the reflection effect (heating by the
radiant energy of the companion), and gravity darkening (variation of
the surface brightness due to the local surface gravity acceleration
change).  For the extrasolar planetary systems, on the other hand, the
radius and mass of a planet are significantly smaller than those of the
host star. Thus, most of those effects can be safely neglected, and one
can derive simpler, and still practically useful, analytic formulae
applying perturbative expansion. In what follows, we present such
analytic expressions for the RM effect with and without the stellar limb
darkening.

\section{Analytic expressions for a uniform stellar disk 
(without limb darkening)}
\label{sec:without_darkening}

\begin{figure}[htb]
\epsscale{0.5}
 \plotone{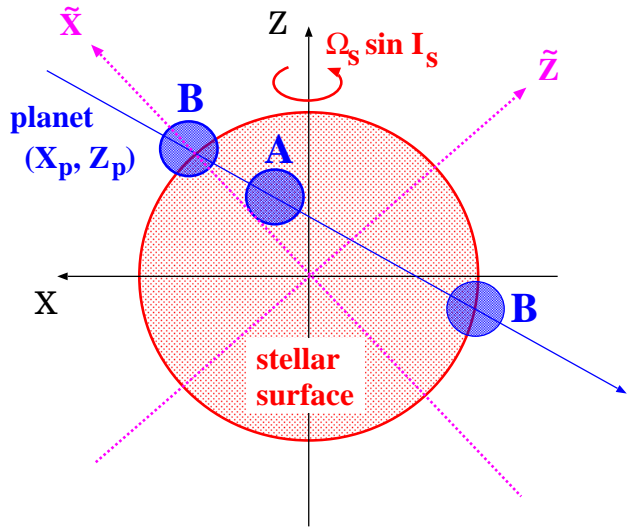}
\caption{Planetary transit: ingress, complete transit, 
and egress phases.
\label{fig:planet-star}}
\end{figure}

As a step toward an analytic model for the RM effect for extrasolar
planetary systems, let us consider first an idealistic case in which the
limb-darkening effect is neglected.  We also assume that the planet is
completely optically thick and not rotating, which is also assumed in
the next section. In this case, one can obtain the exact analytic
expression even without the perturbative expansion.  The intensity at
$(x,z)$ on the uniform stellar surface becomes
\begin{eqnarray}
\label{eq:Ixz}
 I(x,z)=
\cases{I_0 &; $x^2+z^2\leq R_s^2$ and $(x-X_p)^2+(z-Z_p)^2\geq R_p^2$ \cr\cr
       0 &; otherwise \cr
},
\end{eqnarray} 
where $\bm{X}_p=(X_p,Y_p,Z_p)$ is the position of the center of the
planet and $R_s$ and $R_p$ denote the radii of the star and the planet,
respectively. We evaluate equation (\ref{eq:delta_vs}) at complete
transit, ingress, and egress phases in the following subsections (see
Fig.\ref{fig:planet-star}).
%
%
%
%
%
%
\subsection{Complete transit phase} 
\label{subsec:phase_A_no_limb}
%
%
%
%
%
%
During a complete transit phase, the position of the planet satisfies
the relation $\left(X_p^2+Z_p^2\right)^{1/2}<R_s-R_p$.  Thus, the range of the
integral in equation (\ref{eq:delta_vs}) is simply given by the stellar
surface area with the entire planetary disk sutracted, i.e.,
\begin{equation}
\iint dx dz \longrightarrow 
	\int_{-R_s}^{R_s} dx \int_{-\sqrt{R_s^2-x^2}}^{\sqrt{R_s^2-x^2}} dz 
	- \int_{X_p-R_p}^{X_p+R_p} dx 
	\int_{Z_p-\sqrt{R_p^2-(x-X_p)^2}}^{Z_p+\sqrt{R_p^2-(x-X_p)^2}} dz. 
\end{equation}
Then we obtain 
\begin{eqnarray}
 \iint I(x,z)dxdz &=& \pi(R_s^2-R_p^2)I_0 , \\
 \iint xI(x,z)dxdz &=& -X_p \pi R_p^2I_0. 
\end{eqnarray}
Substituting these results into equation (\ref{eq:delta_vs}), we find
\begin{equation}
 \Delta v_{s} = \Omega_s X_p \sin I_s \frac{\gamma^2}{1-\gamma^2} 
\qquad \left( \gamma \equiv \frac{R_p}{R_s} \right) .
\label{eq:v_s_complete}
\end{equation}
Equation (\ref{eq:v_s_complete}) implies that the time dependence of the
RM effect during the complete transit is entirely incorporated in the
planet position, i.e., $X_p=X_p(t)$.

\subsection{Ingress and egress phases} 
\label{subsec:phase_B_no_limb}

\begin{figure}[htb]
\epsscale{0.55} \plotone{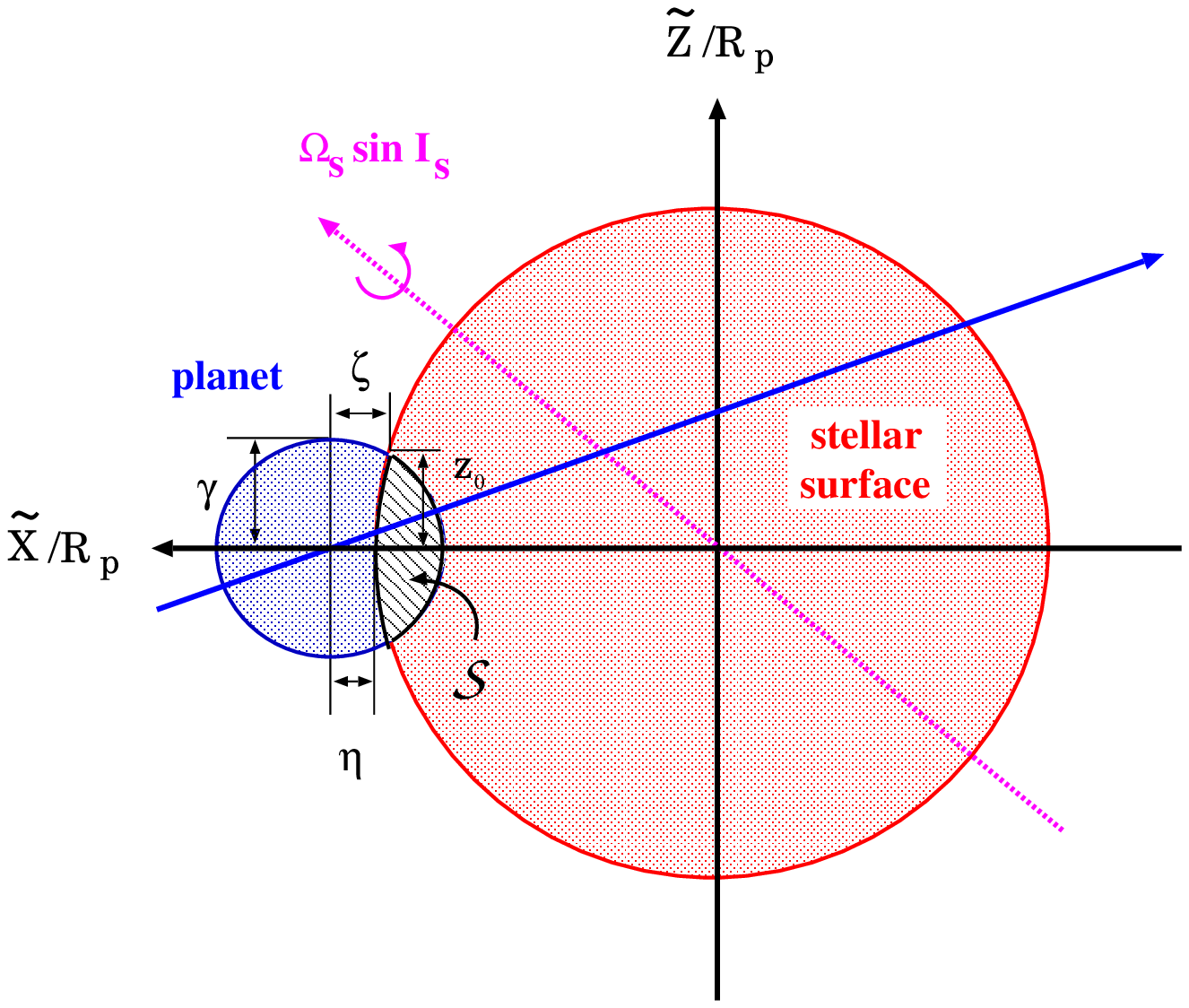} \caption{Schematic
 illustration of the configuration of the system at ingress of the
 planet in the new coordinates.  \label{fig:ingress}}
\end{figure}

At ingress and egress phases, on the other hand, the location of the
planet satisfies the relation, 
$R_s-R_p<\left(X_p^2+Z_p^2\right)^{1/2}<R_s+R_p$.
Just for computational convenience, we rotate the coordinates in a
time-dependent manner so that the planet is always located along the new
$x$-axis:
\begin{eqnarray}
 \left(
\begin{array}{c} 
\tilde{x}\\ 
\tilde{z} 
\end{array}
\right) =  \frac{1}{R_s\sqrt{X_p^2+Z_p^2}} \left(
\begin{array}{cc}
  X_p & Z_p \\
 -Z_p & X_p
\end{array}
\right) \left(
\begin{array}{c} 
x \\ 
z 
\end{array}
\right) .
\end{eqnarray}
Then the position of the planet is given by
\begin{eqnarray}
 \left(
\begin{array}{c} 
\tilde{X_p}\\ 
\tilde{Z_p} 
\end{array}
\right) =  \left(
\begin{array}{c} 
1+ \eta_p \\ 
0
\end{array}
\right) ,
\end{eqnarray}
where
\begin{eqnarray}
\eta_p = \frac{\sqrt{X_p^2+Z_p^2}}{R_s} -1 .
\label{eq:def_eta_p}
\end{eqnarray}
In the new coordinates, equation (\ref{eq:Ixz}) is rewritten as
\begin{eqnarray}
\label{eq:newIxz}
 I(\tilde{x},\tilde{z})
=\cases{
  I_0 &; $\tilde{x}^2+\tilde{z}^2\leq 1$ 
        and $(\tilde{x}-1-\eta_p)^2+\tilde{z}^2\geq \gamma^2$ \cr\cr
  0 &; otherwise
} ,
\end{eqnarray}
and the moments of the intensity reduce to
\begin{eqnarray}
 \iint I(x,z)dxdz &=& R_s^2 \left\{\pi I_0 -
\iint_S I(\tilde{x},\tilde{z})d\tilde{z}d\tilde{x}\right\}, 
\label{eq:moment1} \\
 \iint xI(x,z)dxdz &=& - \frac{R_s^2}{1+\eta_p}
\iint_S (X_p \tilde{x} - Z_p \tilde{z})
I(\tilde{x},\tilde{z})d\tilde{z}d\tilde{x} ,
\label{eq:moment2} 
\end{eqnarray}
where the range of the integrals denoted by $S$ indicates the 
overlapping region between the stellar and the planetary disks 
and can be explicitly written as ({\it dark shaded regions}, 
 Fig. \ref{fig:ingress}): 
\begin{eqnarray}
\iint_S d\tilde{z}d\tilde{x}~~~ \longrightarrow  ~~~
\int_{x_0}^1 d\tilde{x}\int_{-\sqrt{1-\tilde{x}^2}}^{\sqrt{1-\tilde{x}^2}}
d\tilde{z}
\,\,+\,\, \int_{\tilde{X}_p-\gamma}^{x_0} d\tilde{x} 
\int_{-\sqrt{\gamma^2-(\tilde{x}-\tilde{X}_p)^2}}^{\sqrt{\gamma^2-
(\tilde{x}-\tilde{X}_p)^2}}d\tilde{z}.
\label{eq:range_of_int}
\end{eqnarray}
Note that the planetary and stellar circles intersect at $(x_0,\pm
z_0)$, where
\begin{eqnarray}
\label{eq:x0z0}
 x_0=1-\frac{\gamma^2-\eta_p^2}{2(1+\eta_p)},
\quad ~
z_0= \sqrt{1-x_0^2} 
= \frac{\sqrt{(\gamma^2-\eta_p^2)[(\eta_p+2)^2-\gamma^2]}}
{2(1+\eta_p)}.
\end{eqnarray}
Let us also introduce
\begin{eqnarray}
\label{eq:zeta}
\zeta \equiv 1+\eta_p-x_0
= \frac{2\eta_p + \gamma^2 + \eta_p^2}{2(1+\eta_p)}.
\end{eqnarray}
Physically speaking, this corresponds to the separation between the
intersection and the center of the planet along the $\tilde{X}$ axis,
but we allow $\zeta$ to be negative (see Fig. \ref{fig:ingress}) as
well.  Then equations (\ref{eq:moment1}) and (\ref{eq:moment2}) are
analytically integrated as
\begin{eqnarray}
&& \iint_S I(\tilde{x},\tilde{z})d\tilde{z}d\tilde{x} 
= I_0\left[\,\,\sin^{-1}z_0 - (1+\eta_p)z_0
+\gamma^2\cos^{-1}(\zeta/\gamma)\,\,\right], 
\label{eq:integral1}
\end{eqnarray}
and 
\begin{eqnarray}
&& \iint_S (\tilde{x}X_p-\tilde{z}Z_p)I(\tilde{x},\tilde{z})
d\tilde{z}d\tilde{x}=
 I_0X_p(1+\eta_p)\left[-z_0\zeta+\gamma^2\cos^{-1}(\zeta/\gamma)\right], 
\label{eq:integral2}
\end{eqnarray}
respectively.

Combining these results, we finally find that equation
(\ref{eq:delta_vs}) reduces to
\begin{eqnarray}
\label{eq:v_s_ingree/egress}
\Delta v_{\rm s}=  \Omega_s X_p \sin I_s \,\,
\frac{-z_0\zeta + \gamma^2\cos^{-1}(\zeta/\gamma)}
{\pi-\sin^{-1}z_0+(1+\eta_p)z_0
-\gamma^2\cos^{-1}(\zeta/\gamma)} .
\end{eqnarray}
While the above expression for $\Delta v_s$ seems a bit complicated,
this is the exact result for the radial velocity anomaly
through relations (\ref{eq:def_eta_p}), (\ref{eq:x0z0}) and
(\ref{eq:zeta}) in terms of the planet position $(X_p,~Z_p)$ specified
by equation (\ref{eq:planet_position}).
%
%
%
%
%
%
%
%
%
%
\section{Effect of stellar limb darkening}
\label{sec:limb_darkening}
%
%
%
%
%
%
To be more realistic, we now take account of the effect of limb
darkening, which produces the radial dependence of the intensity of the
stellar disk. Among the various models proposed so far
\citep[e.g.,][]{Claret00}, we adopt a linear limb-darkening law as the
simplest, but a practically realistic one.  Introducing the linear
limb-darkening coefficient $\epsilon$, the stellar intensity is now
given by
\begin{eqnarray}
\label{eq:I_linearlimb}
I(x,z)= \left\{
\begin{array}{l}
 I_0 \,\,\{1-\epsilon(1-\mu)\,\,\} ~~
;~~x^2+z^2\leq R_s^2~~ \mbox{and}~~ (x-X_p)^2+(z-Z_p)^2\geq R_p^2 
\\ 
\\
 0 ~~~~; ~~\mbox{otherwise}
\end{array}
\right. ,
\label{eq:I_limb_darkening}
\end{eqnarray}
where $\mu$ is the cosine of the angle between the line of sight and the 
vector normal to the local stellar surface : 
\begin{equation}
\mu = \sqrt{1-\frac{x^2+z^2}{R_s^2}}.
\end{equation}
With the limb-darkening effect, however, equation (\ref{eq:delta_vs})
can no longer be analytically integrated in an exact manner. Therefore,
we construct approximate analytic formulae on the basis of the result
without limb darkening ($\epsilon=0$; see section
\ref{sec:without_darkening}).
%
%
%
%
%
%
\subsection{Complete transit phase} 
\label{subsec:phase_A_limb}
\begin{figure}[t]
 \epsscale{0.9}
 \plottwo{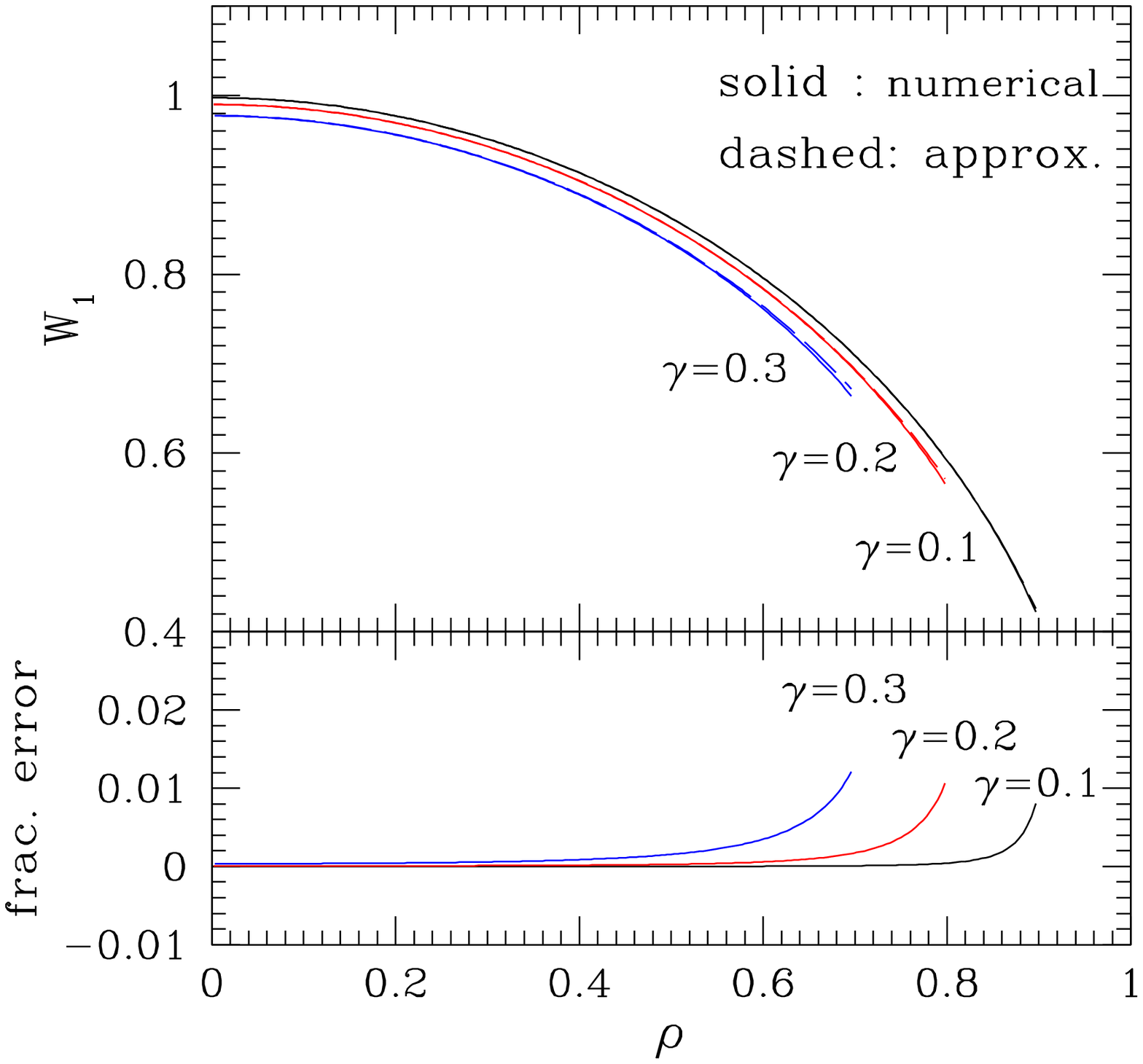}{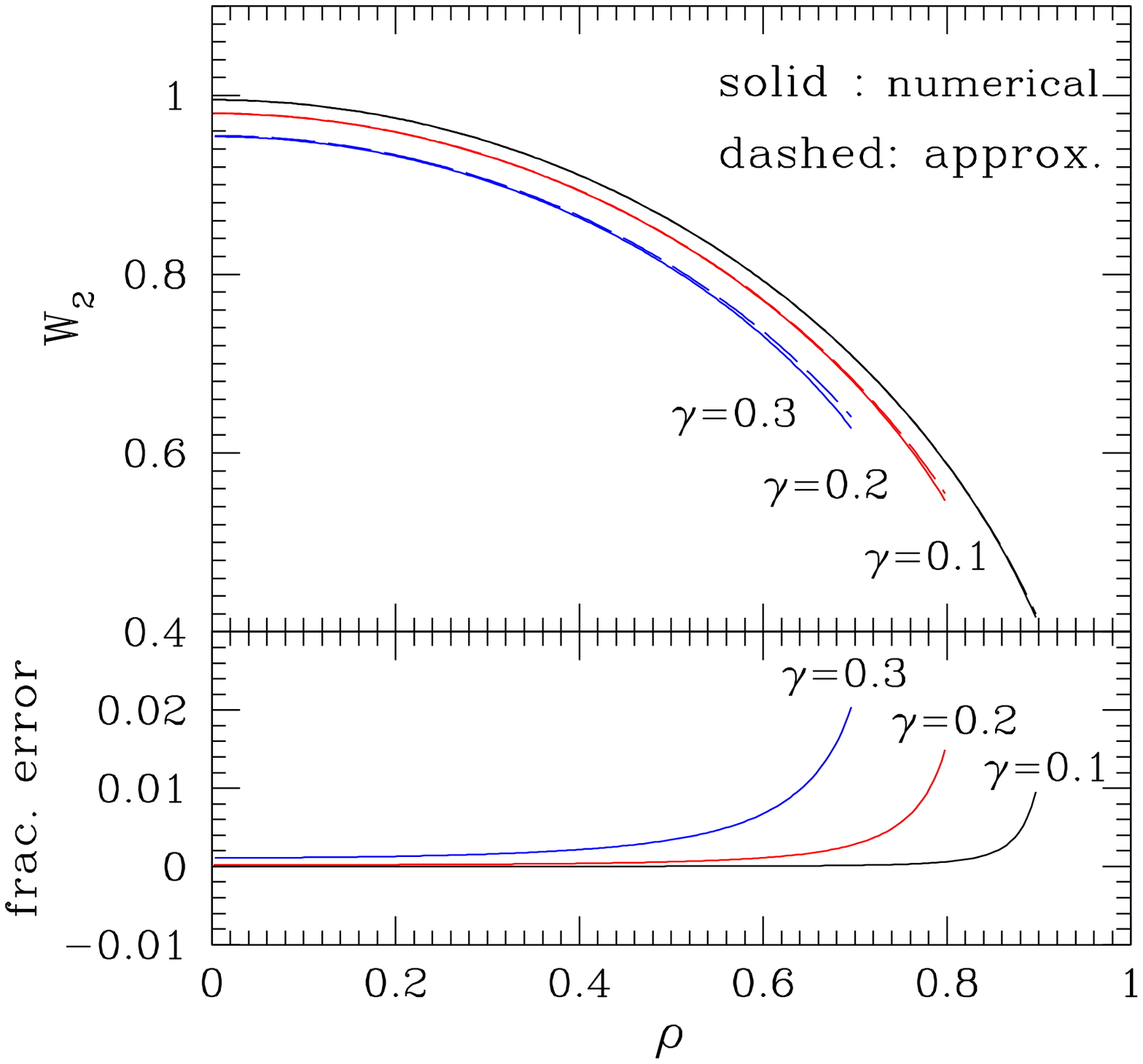}
 \figcaption{{\it Top windows}: Approximate vs. numerical evaluations of
 the integrals $W_1$ ({\it left}) and $W_2$ ({\it right}) 
 as a function of $\rho=\left(X_p^2+Z_p^2\right)^{1/2}/R_s$. The solid lines 
 represent the numerical evaluation of equations 
 (\ref{appen:W_1_reduced}) and (\ref{eq:W_2_reduced}), 
 while the dashed lines are the approximation based on the perturbative 
 expansions (\ref{appen:approx_W_1}) and (\ref{appen:approx_W_2}). 
 Note that the variable $\rho$ runs from $0$ to $1-\gamma$. 
 {\it Bottom windows}: Fractional errors $W^{(\rm approx)}/W^{\rm (num)}-1$ 
 for the integrals $W_1$ ({\it left}) and  $W_2$ ({\it right}). 
 \label{fig:integral_W1_W2}} 
\end{figure}

%
%
%
%
%
%
Applying the analytic results of \S\ref{subsec:phase_A_no_limb} to the
stellar intensity model (\ref{eq:I_linearlimb}), 
equation (\ref{eq:delta_vs}) is formally rewritten as
\begin{eqnarray}
\label{eq:dV_s_limbdarkening_A}
\Delta v_{\rm s}=  \Omega_s X_p \sin I_s \,\,
\frac{ \gamma^2\,\{1-\epsilon(1-W_2)\}}
{ 1-\gamma^2-\epsilon\left\{\frac{1}{3}-\gamma^2(1-W_1)\right\}} ,
\end{eqnarray}
where $W_1$  and $W_2$ are defined as
\begin{eqnarray}
W_1&=& \frac{1}{\pi R_p^2}\, \iint_S dxdz \sqrt{1-(x^2+z^2)/R_s^2},
\label{eq:W_1}
\\
W_2&=& \frac{1}{X_p\, \pi R_p^2}\,
\iint_S dx dz\,\, x\sqrt{1-(x^2+z^2)/R_s^2}.
\label{eq:W_2}
\end{eqnarray}
The above integrals are carried out over the entire planetary disk.  As
discussed in Appendix \ref{subsec:W_1_2}, they reduce to one-dimensional
integrals, which can be expanded with respect to $\gamma =
R_p/R_s$. Specifically, equations (\ref{appen:approx_W_1}) and
(\ref{appen:approx_W_2}) show perturbative expressions up to the fourth
order in $\gamma$.  The accuracy of the fourth-order perturbation
expansion is within a few percent even for $\gamma\sim0.3$
(Fig.\ref{fig:integral_W1_W2}). In practice, however, the value of
$\gamma$ is expected to be much smaller, $\gamma\simlt0.1$.  In this
case, higher order terms in equations (\ref{appen:approx_W_1}) and
(\ref{appen:approx_W_2}) contribute merely $\sim 1$\% to equation
(\ref{eq:dV_s_limbdarkening_A}), and one can safely use
\begin{eqnarray}
W_1(\rho) & \simeq & 0, 
\label{eq:approx_W1} \\
W_2(\rho) & \simeq & (1-\rho^2)^{1/2},  
\label{eq:approx_W2}
\end{eqnarray}
where  $\rho \equiv \left(X_p^2+Z_p^2\right)^{1/2}/R_s$ ($0<\rho<1-\gamma$). 

\subsection{Ingress and egress phases} 
\label{subsec:phase_B_limb}

If the linear limb-darkening effect is taken into account, equation
(\ref{eq:v_s_ingree/egress}), describing the ingress and egress phases,
now becomes
\begin{eqnarray}
\Delta v_{\rm s} &=& \Omega_s X_p \sin I_s \cr
&&\times\,\,
\frac{\displaystyle 
(1-\epsilon)\left\{-z_0\zeta + \gamma^2\cos^{-1}(\zeta/\gamma)\right\} 
+ \frac{\epsilon}{1+\eta_p}\,\, W_4}
{\pi\left(1-\frac{1}{3}\epsilon\right)- (1-\epsilon)\left\{\sin^{-1}z_0-(1+\eta_p)z_0
+\gamma^2\cos^{-1}(\zeta/\gamma)\right\}-\epsilon\,\,W_3},
\label{eq:dV_s_limbdarkening_B}
\end{eqnarray}
where  $W_3$ and $W_4$ are defined  by 
\begin{eqnarray}
W_3 &=& \iint_S d\tilde{x}d\tilde{z} \sqrt{1-\tilde{x}^2-\tilde{z}^2},
\label{eq:W_3} \\
W_4 &=& \iint_S d\tilde{x}d\tilde{z} \,\,\tilde{x}
\sqrt{1-\tilde{x}^2-\tilde{z}^2}.
\label{eq:W_4} 
\end{eqnarray}

Appendix \ref{subsec:W_3_4} derives approximate analytic expressions
(\ref{appen:approx_W_3}) and (\ref{appen:approx_W_4}) for equations
(\ref{eq:W_3}) and (\ref{eq:W_4}), respectively, assuming that $\gamma
\ll 1$. Again, if $\gamma\simlt0.1$, they can be safely set as
\begin{eqnarray}
W_3&\simeq& 0,  
\label{eq:approx_W3} \\
W_4&\simeq& \frac{\pi}{2}\,\,\gamma(\gamma-\zeta)\,\,x_c
\,\,\frac{g(x_c\,;\,\eta_p,\,\gamma)}
{g(1-\gamma;\,-\gamma,\,\gamma)}\,\,W_2(1-\gamma),
\label{eq:approx_W4} 
\end{eqnarray}
where
\begin{eqnarray}
x_c &=& x_0 + \frac{\zeta-\gamma}{2}, \\
g(x;\,\eta_p,\,\gamma) &=& (1-x^2)\sin^{-1}
\left\{\frac{\gamma^2-(x-1-\eta_p)^2}{1-x^2}\right\}^{1/2} \cr
&& ~~~~~~ + \sqrt{\{\gamma^2-(x-1-\eta_p)^2\}
	\{1-x^2-\gamma^2+(x-1-\eta_p)^2\}}.
\end{eqnarray}
As shown in Figure \ref{fig:integral_W3_W4}, the accuracy of equations
(\ref{eq:approx_W3}) and (\ref{eq:approx_W4}) is typically within a
fractional error of $5\%$-$10\%$ percent. Nevertheless, their contribution
to the total error budget for the velocity anomaly
(\ref{eq:dV_s_limbdarkening_B}) is within a few percent (see
\S\ref{subsec:num_accuracy}). Thus, equations (\ref{eq:approx_W3}) and
(\ref{eq:approx_W4}) are practically good approximations in most cases.

\section{Application to the HD 209458 system \label{sec:HD209458}}

So far, HD 209458 is the only extrasolar planetary system in which the RM
effect is detected; \citet{Queloz00} reported the first detection of
this effect with ELODIE spectrograph on the 193 cm telescope of the
Observatoire de Haute Provence. They numerically computed the radial
velocity anomaly due to the RM effect for a variety of model parameters
and compared these with the observed radial curves. They concluded that
$\alpha=\pm3.\!\!^\circ 9 {+18^\circ \atop -21^\circ}$ and $V \sin I_s=3.75\pm 1.25$~km
s$^{-1}$, where $\alpha$ is the angle between the planet's orbital plane
and the star's apparent equatorial plane and $V$ denotes the stellar
surface velocity.  These are written as $\alpha =\cos^{-1}(\sin
i\cos\lambda)$ and $V=R_s\Omega_s$ according to the notation of our
current paper.  We summarize the current estimates of the stellar and
planetary parameters for HD 209458 in Table \ref{tab:planetsystem} and
the best solution for the spin parameters by \citet{Queloz00} in Table
\ref{tab:diff_notation}.  Since HD 209458 remains
the best target for the precise measurement of the RM effect,
we consider in this section the extent to which one can improve the
constraints on the spin parameters with our analytic formulae.

\subsection{Parameter dependence} 
\label{subsec:param_depend}

Adopting the linear limb-darkening law for the stellar intensity model,
the RM effect for a system in the Keplerian orbit is specified by $10$
parameters: the limb-darkening coefficient $\epsilon$, the orbital
parameters of the system ($a$, $e$, $i$, $\varpi$, and $P_{\rm orb}$),
the size of
the stellar and planetary disks ($R_s$ and $R_p$), the projected stellar
surface velocity $V\sin I_s$, and the projected angle between
the stellar spin axis and the normal direction of the orbital plane
$\lambda$. Except for the last two parameters ($V\sin I_s$ and $\lambda$),
these can be independently determined from the
usual radial velocity and transiting photometric data, at least in
principle.  This is indeed the case for the HD 209458 system (see Table
\ref{tab:planetsystem}). Therefore, it is natural to ask about the extent to
which one can put constraints on the two parameters $V\sin I_s$ and
$\lambda$ from the radial velocity anomaly during the transit due to the
RM effect.

\begin{figure}[htb]
 \epsscale{0.55}
 \plotone{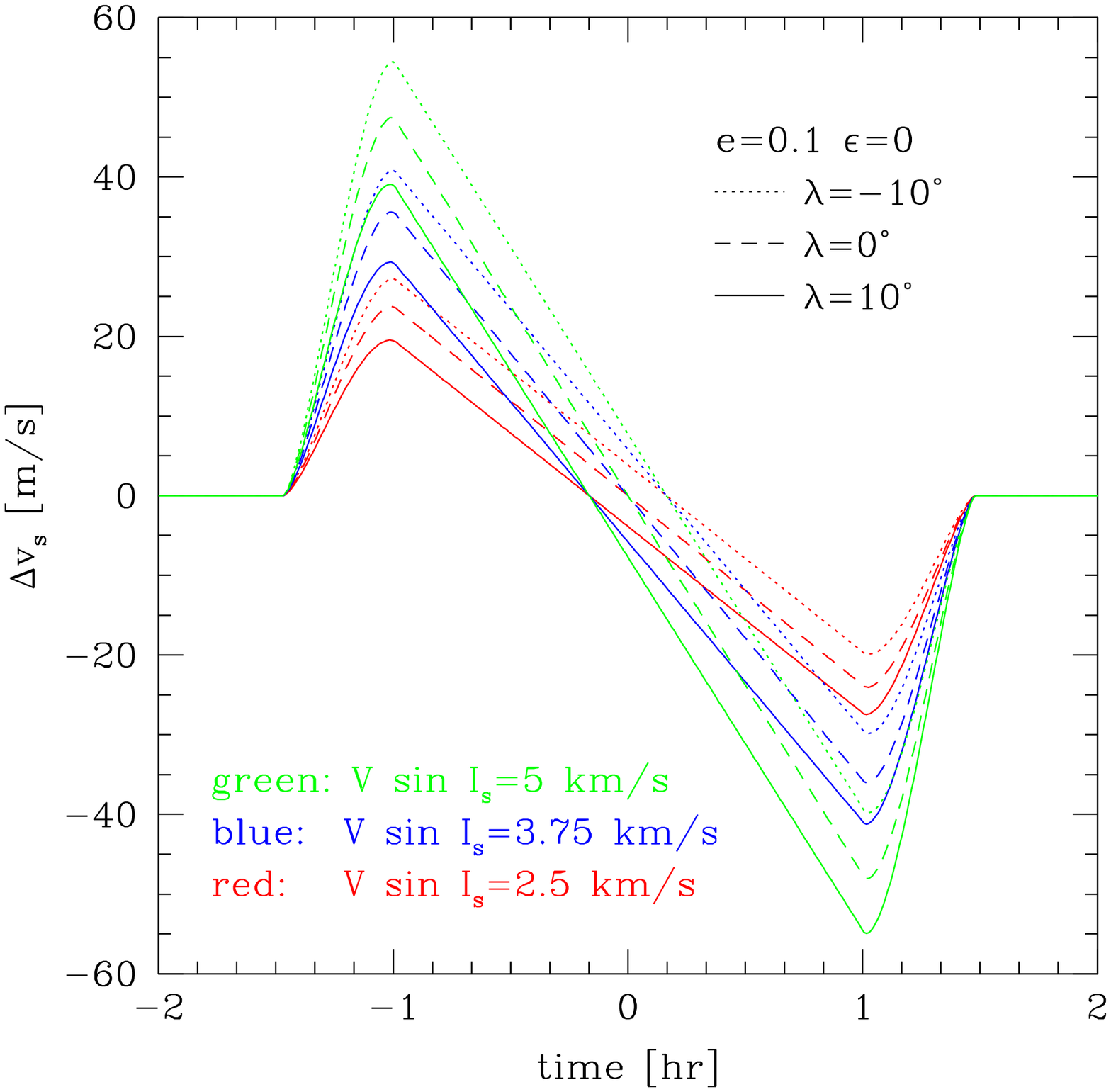}
 \figcaption{Analytic radial velocity curves for the RM effect 
	without stellar limb darkening ($\epsilon=0$).
 \label{fig:radial_veloc_no_limb}}

\vspace*{0.5cm}

 \epsscale{0.55}
 \plotone{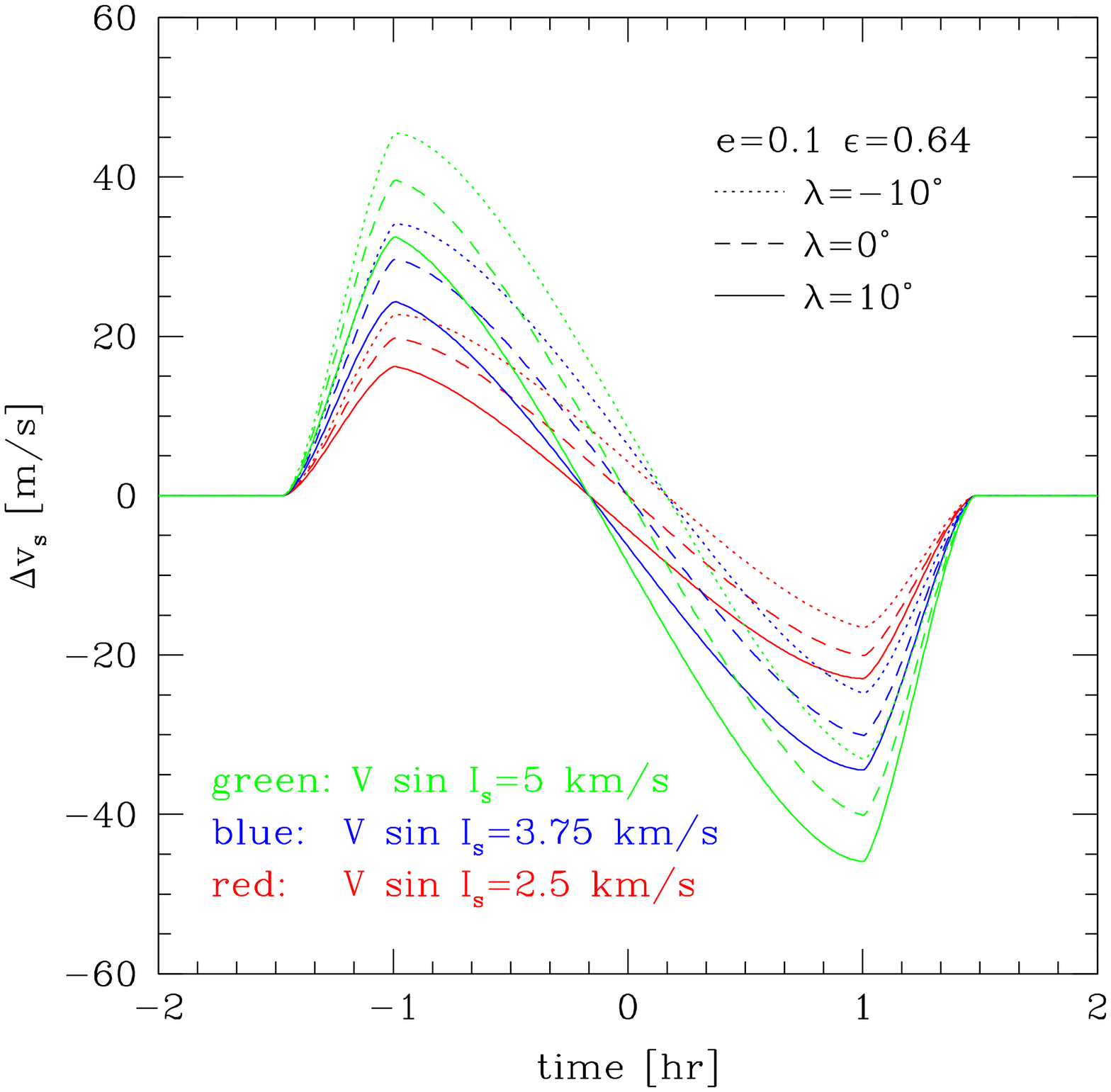}
 \figcaption{Same as Fig.\ref{fig:radial_veloc_no_limb}, but
with a linear limb-darkening effect ($\epsilon=0.64$).
 \label{fig:radial_veloc_limb064}}
\end{figure}

Consider first the sensitivity to the spin parameters ($V\sin I_s$,
$\lambda$).  Figures \ref{fig:radial_veloc_no_limb} and
\ref{fig:radial_veloc_limb064} illustrate our approximations for
$\Delta v_s$ adopting the estimated parameters of the HD 209458 system
(Table \ref{tab:planetsystem}) with and without the stellar limb
darkening (i.e., $\epsilon=0$ and $0.64$, specifically),
respectively. The central transit epoch is chosen as $t=0$. Then ingress
starts at $t=-1.55$ hr, the complete transit lasts for $-1.07<t<1.08$ hr,
and egress ends at $t=1.56$ hr for $e=0.1$ (sometimes these
four epochs are referred to as the first, second, third, and fourth
contacts, respectively).  The range of the spin parameters, $V\sin I_s$
and $\lambda$, adopted in these figures roughly covers the uncertainties
of the values of \citet{Queloz00}.

Comparison of the two figures indicates that the radial velocity anomaly
$\Delta v_s$ is also sensitive to the linear limb-darkening coefficient
$\epsilon$. Obviously the amplitude of the radial velocity shift $\Delta
v_r$ is sensitive to $V\sin I_s$. The projected angle $\lambda$ shifts
the zero point of the radial velocity anomaly at earlier
($\lambda>0$) and later ($\lambda<0$) epochs for the 
orbital inclination $i<90^{\circ}$. 
This produces an asymmetry of the shape of the radial velocity 
anomaly. 
Note that the behavior becomes opposite for the 
inclination $i>90^{\circ}$, corresponding to the parameter degeneracy 
between $(i,\lambda)$ and $(180^{\circ}-i, -\lambda)$.
Because of the different dependence of the overall radial velocity anomaly 
on the spin parameters $(V\sin I_s,~\lambda)$, one can put more stringent
constraints on those if our formulae are combined with future precision
data attainable by 8 -- 10 m class telescopes with a high dispersion
spectrograph (HDS) such as Subaru HDS.

\begin{figure}[htb]
 \epsscale{1.0} \plotone{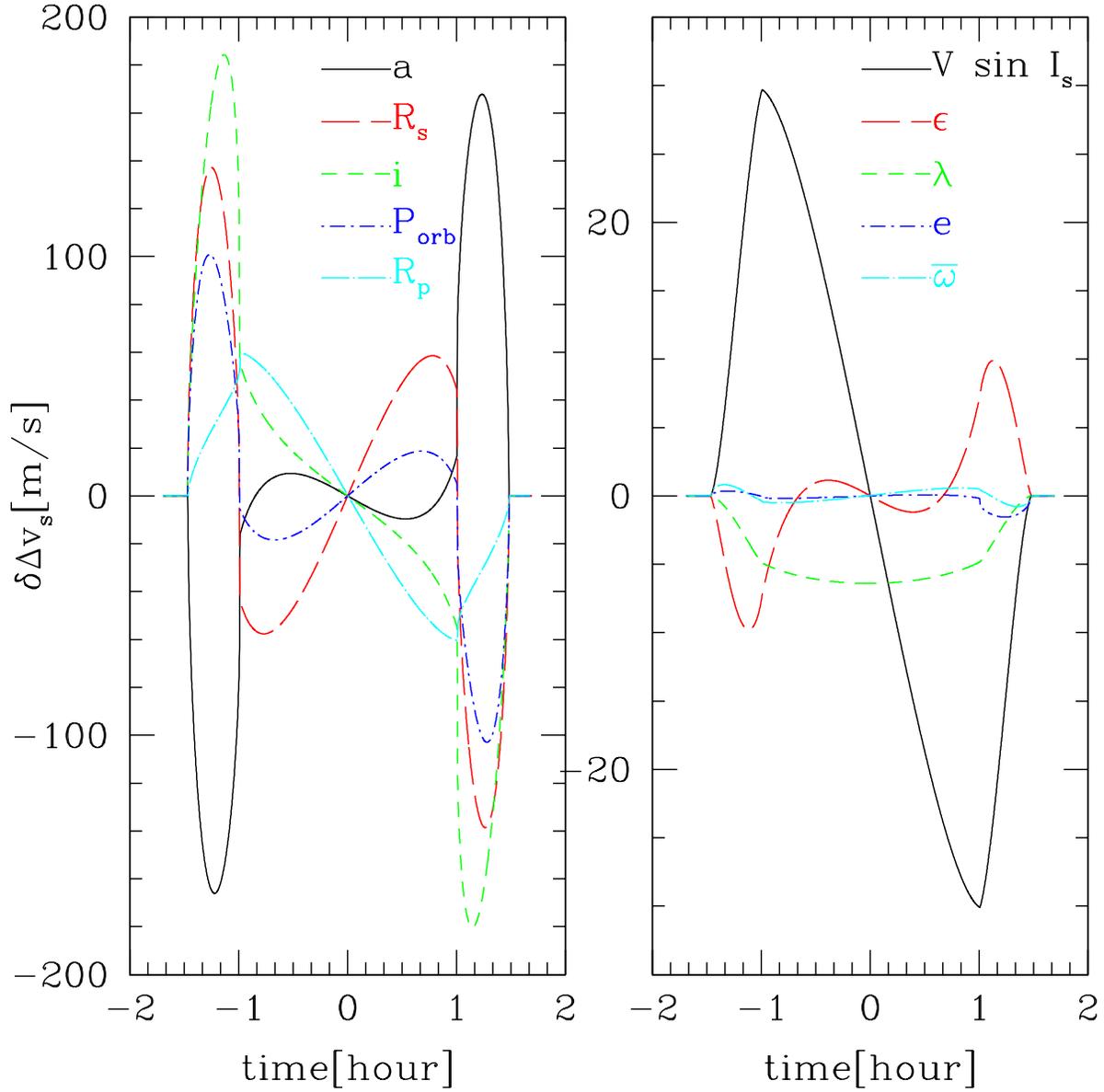} \figcaption{Variation of radial
velocity shift with respect to a parameter variation $p\to p+dp$ as a
function of time.  The vertical axis denotes the variation of radial
velocity shift normalized by the fractional error in each parameter
$dp/p$ ( eqs. [\ref{eq:errors_in_delta_v_s_1}] and
[\ref{eq:errors_in_delta_v_s_2}] for left and right panels,
respectively).  These quantities are evaluated around the fiducial
values for the HD 209458 system summarized in Table
\ref{tab:planetsystem} using the analytic formulae for the RM effect.
\label{fig:parameter_dependence} }
\end{figure}

Before addressing this issue in detail, it is helpful to clarify
the dependence of the RM effect on the other
remaining parameters. To investigate this, we quantify the variation of
the radial velocity shift with respect to a specific parameter change
$p\sim p+dp$ by
\begin{equation}
 \delta\Delta v_s\equiv \lim_{dp\to0} 
\frac{\Delta v_s(p+dp)-\Delta v_s(p)}{dp/p}
\label{eq:errors_in_delta_v_s_1}
\end{equation}
for $p=a, R_s,R_p,P_{\rm orb},V \sin I_s,\epsilon,$ and $e$.  In
practice, we systematically decrease the value of $dp/p$ for each
parameter down to $10^{-6}$ and ensure the convergence of the
derivative.  For the angular parameters $p=i,\varpi,$ and $\lambda$, we
simply take their scaling values at $10^\circ$:
\begin{equation}
 \delta\Delta v_s\equiv \lim_{dp\to0} 
\frac{\Delta v_s(p+dp)-\Delta v_s(p)}{dp/10^\circ} .
\label{eq:errors_in_delta_v_s_2}
\end{equation}
Here, we confirm the convergence of the derivative by decreasing the
value of $dp/10^\circ$ down to $10^{-6}$ in these cases.  Our analytic
formulae are indeed useful in evaluating these quantities at the
fiducial parameters of the HD 209458 system (Tables
\ref{tab:planetsystem} and \ref{tab:diff_notation}). The results are
plotted in Figure \ref{fig:parameter_dependence} as a function of time.
Note that our definition of $\delta\Delta v_s$ is {\it normalized}
by the fractional error in the parameter, i.e., $dp/p$.

Figure \ref{fig:parameter_dependence} clearly shows that the stellar
radius $R_s$ and the orbital parameters $a$, $i$, and $P_{\rm orb}$
sensitively change the normalized radial velocity variation
$\delta\Delta v_s$ at the ingress and egress phases, while the spin
parameters $(V\sin I_s, \lambda)$ have a relatively smaller effect on
$\delta\Delta v_s$.  To quantify the actual deviation of the radial
velocity shift caused by the systematic errors in the observation, we must
multiply the observational uncertainty listed in Table
\ref{tab:planetsystem}.  Then it turns out that the most sensitive
parameter is $V\sin I_s$, causing the $12$m s$^{-1}$ variation of the
radial velocity shift. The other parameters change the radial
velocity shift at the ingress and egress phases by less than $5$m s$^{-1}$.
This amplitude itself is comparable to the $\delta\Delta v_s$ induced by the
uncertainty of the projected angle $\lambda$.  Nevertheless, the
different time-dependent effects of the various parameters can be used to
break the parameter degeneracy, which would enable an accurate
determination of the spin parameters $\lambda$ and $V\sin I_s$. In
addition, Figure \ref{fig:parameter_dependence} even suggests that a
more precise determination of the orbital parameters other than the spin
parameters is also possible by combining the RM effect with the usual radial
velocity measurement.

\subsection{Accuracy of our formulae} 
\label{subsec:num_accuracy}

\begin{figure}[htb]
 \epsscale{0.49} \plotone{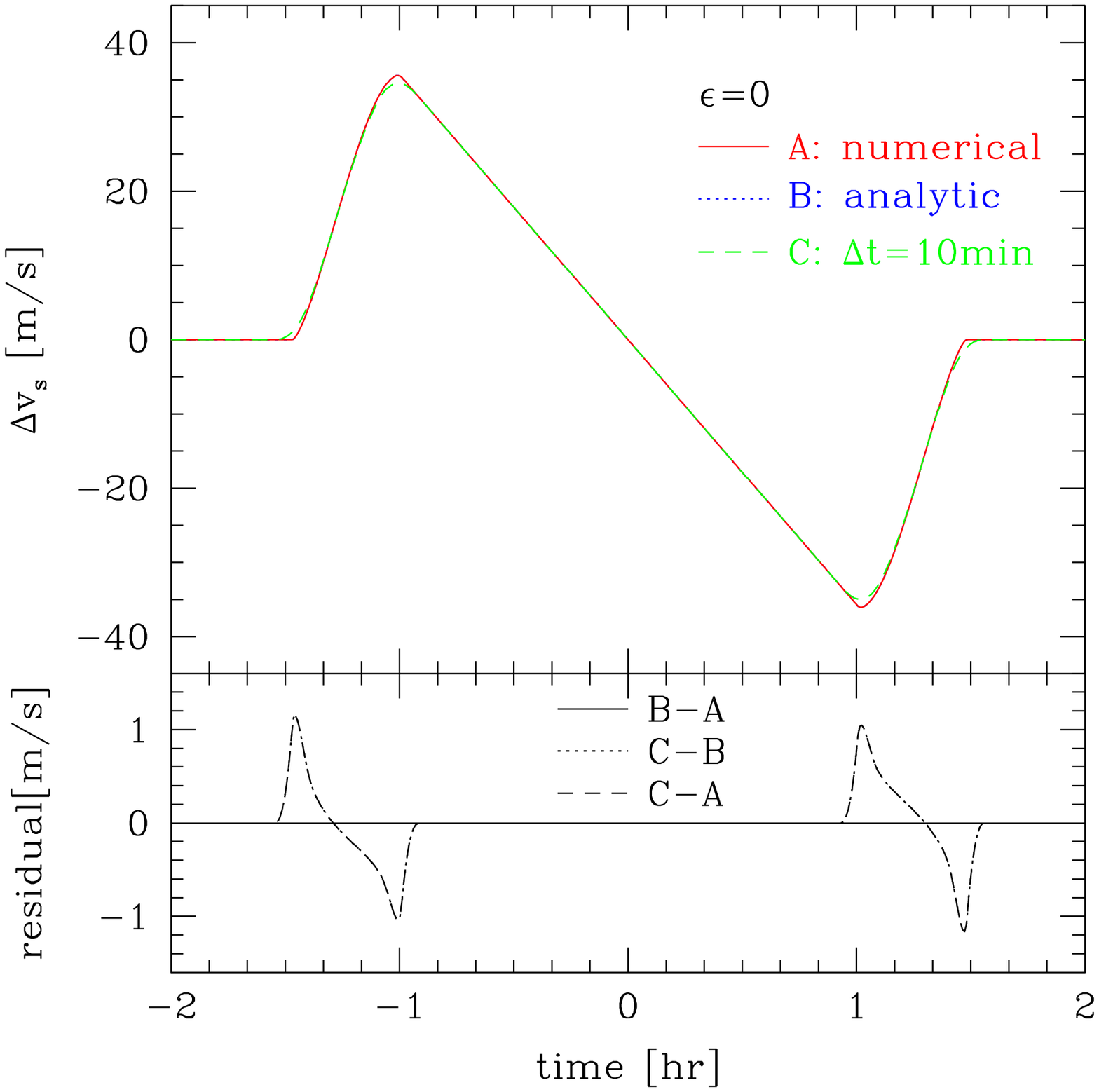} 
\figcaption{Comparison of different models of the radial velocity
 anomaly curves for the HD 209458 system without limb darkening
 ($\epsilon=0$). The top panel shows the radial velocity shifts
 obtained from the numerical evaluation of expression
 (\ref{eq:delta_vs}) ({\it solid curve}) and the analytic formulae with
 and without the effect of exposure time ({\it dotted and
 dashed curves, respectively}). The bottom panel shows the residuals
 between the two radial  velocity shifts among three curves.
 \label{fig:differences_eps0}}
\vspace*{0.6cm}

 \epsscale{0.49}
 \plotone{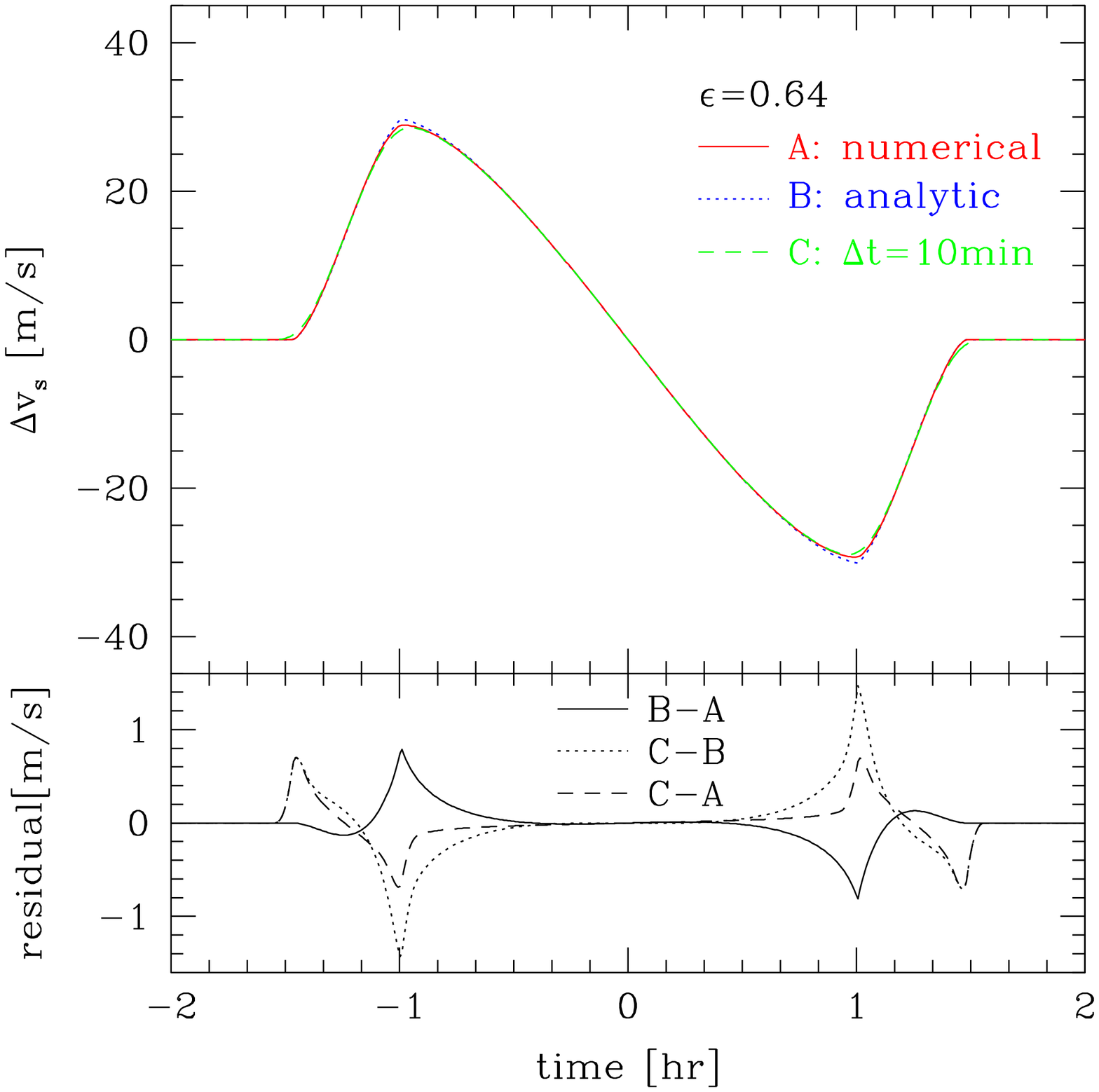}
 \figcaption{Same as Fig.\ref{fig:differences_eps0}, but with limb
 darkening ($\epsilon=0.64$).
 \label{fig:differences_eps0.64}}
\end{figure}

While our analytic formulae presented in the previous section will
improve the efficiency of the parameter estimations relative to fully
numerical approaches, we have to address a couple of issues before
applying them to real data: their accuracy and the effect of the finite
exposure time.  Our formulae with limb darkening are derived on the
basis of an empirical approximation to the integrals of the stellar
surface intensity (\S \ref{sec:limb_darkening}). Furthermore, the real
data do not instantaneously sample the radial velocity, but are averaged
over a finite exposure time.  We directly test those effects
against the numerical solutions of equation (\ref{eq:delta_vs}).
Figures \ref{fig:differences_eps0} and \ref{fig:differences_eps0.64}
compare three results: numerical integration of equation
(\ref{eq:delta_vs}), our analytic formulae, and the numerical average of
equation (\ref{eq:delta_vs}) over a $\Delta t=10$ minute exposure time (in
practice, we separately average the denominator and the numerator of the
analytic formulae assuming $\Delta t=10$ minutes and then take their
ratio). These results are labeled A, B, and C and plotted in solid,
dotted, and dashed curves, respectively, in the top panels.

For $\epsilon=0$, the analytic formulae (curve B) are exact, and the
completely negligible difference A$-$B should be regarded as a welcome
check of the accuracy of our numerical integration scheme. The bottom
panels suggest that the three results agree within an accuracy of $\sim
1$m s$^{-1}$, which is below the typical radial velocity sensitivity
achieved ($\sim 3$m s$^{-1}$) and is only comparable to the latest
achievement by HARPS \citep{HARPS}. Therefore, as far as the HD 209458
system is concerned, we can safely use our analytic formulae as useful
templates for the RM effect even if a finite exposure time of the order
of 10 minutes is taken into account.

\subsection{Mock analysis of the spin parameter estimation} 
\label{subsec:param_estimation}

Now we are in a position to ask whether our formulae combined with precision
spectroscopic data can improve the previous constraints on the spin
parameters $(\Omega_s\sin I_s,\,\lambda)$, among others.  For this
purpose, we create mock data for the radial velocity anomaly of the
HD 209458 system and fit them to the analytic formulae.  Basically, the
mock data were created adopting the central values of the
parameters listed in Tables \ref{tab:planetsystem} and
\ref{tab:diff_notation}, but assigned an overall Gaussian random error
of the rms amplitude $5$ m s$^{-1}$, which is the level of precision
achieved with the Subaru HDS assuming an
exposure time of $\Delta t=10$ minutes \citep[][]{Winn2004}.  In
light of the most recent sensitivity achieved by HARPS
(\citealt{HARPS}; $\sim 1$m s$^{-1}$), our error assignment may still be
conservative if
the error is not dominated by other possible systematic errors.  To mimic the
effect of the finite exposure time, we numerically integrate the
denominator and the numerator of equation (\ref{eq:delta_vs}) separately
over $\Delta t=10$ minutes. Then we take the ratio and assign the random
error, as mentioned above.

Note that the number of independent data points during the transit phase
($\sim 3$ hr including the ingress and egress phases) is $16$ for
$\Delta t=10$ minutes.  The generated mock data are then fitted to the
analytic radial velocity anomaly to estimate the spin parameters. Here,
the fitting is performed assuming prior knowledge of the remaining
eight parameters.

\begin{figure}[htb]
 \epsscale{0.85} \plotone{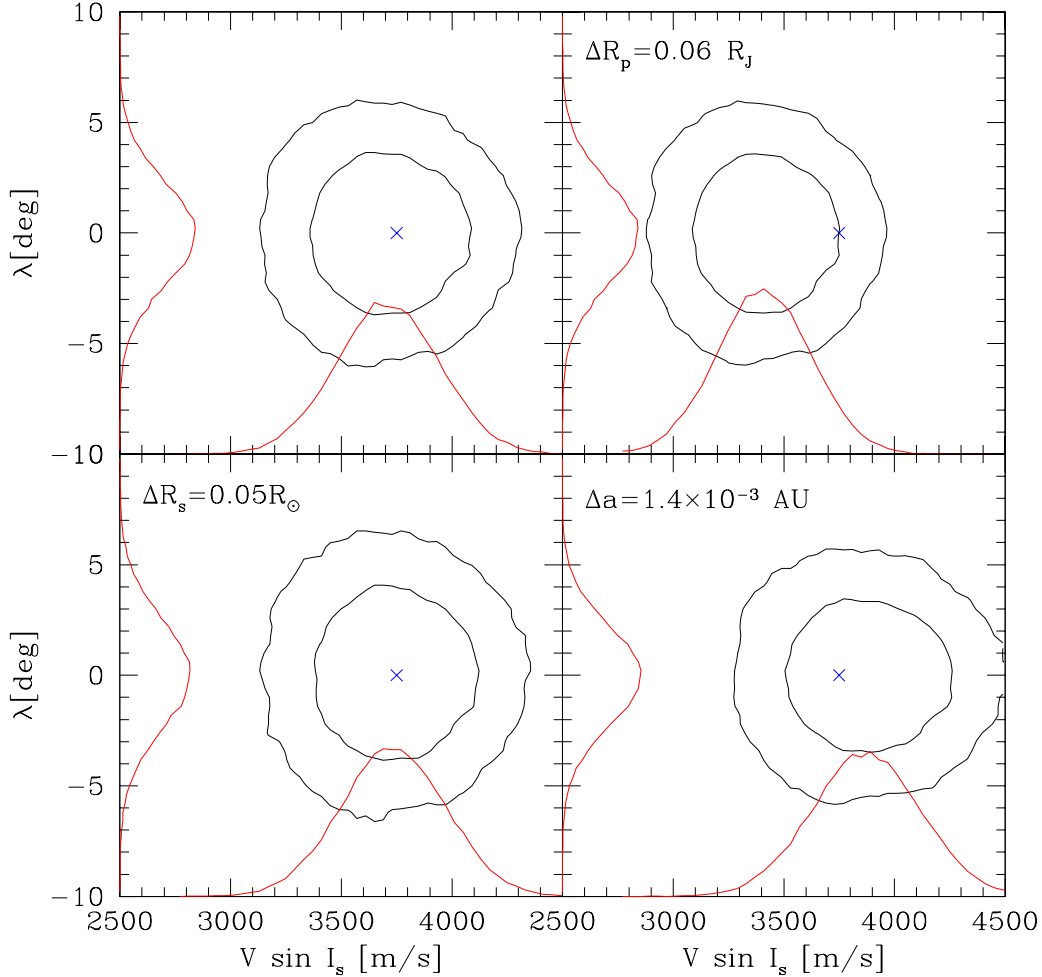} 

\figcaption{Joint probability distribution of the best-fit parameters
 $(V\sin I_s,~\lambda)$ in the likelihood analysis for the mock HD 209458
 observation.  We adopt the prior assumption for fiducial parameters
 ({\it top left}), $R_p=1.407 \,R_{\rm J}$ ({\it top right}),
 $R_s=1.196 \,R_\odot$ ({\it bottom left}) and $a=0.0482\,{\rm
 AU}$ ({\it bottom right}). In each panel, the cross indicates the
 correct values of $(V\sin I_s,~\lambda)$, and the contour levels around
 it represent the 1 and 2 $\sigma$ levels of the probability
 distribution. The solid lines projected onto the horizontal and vertical
 axes represent the probability distributions of $V\sin I_s$ and
 $\lambda$, respectively.
 Note that all the mock data created in the likelihood analysis
 assume the parameters listed in Table \ref{tab:planetsystem}.  Thus,
 except for the top left panel, the difference between the incorrect prior
 assumption and the correct value for $(R_p,R_s,a)$ was used in each panel.
 \label{fig:probdist_best-fit}}
\end{figure}

First, let us see how the spin parameters are reliably estimated from the
$\chi^2$ fit.  To examine this, we create $50,000$ mock
realizations and calculate the joint probability distribution of the
best-fit parameters under a certain prior knowledge of $R_s$, $R_p$, and
$a$.  We use the $\chi^2$ function,
\begin{equation}
\chi^2(V\sin I_s, \lambda) 
= \sum_{i=1}^{N}\left\{\frac{\Delta v_{s,i}^{\rm data}-\Delta v_s^{\rm model}
(V\sin I_s, \lambda;t_i)}
{\sigma_{\scriptscriptstyle \Delta v}}\right\}^2
\end{equation}
with $\sigma_{\scriptscriptstyle\Delta v}=5$m s$^{-1}$ and $N=16$.  In
this analysis, according to the result in Figure
\ref{fig:parameter_dependence}, we particularly focus on the five
parameters $V \sin I_s$, $\lambda$, $R_s$, $R_p$, and $a$. Their input
values are $3750$ m s${}^{-1}$, $0^\circ$, $1.146{\rm R_\odot}$,
$1.347{\rm R_J}$, and $0.0468$ AU, respectively.  We assume a set of
different prior values for $R_s$, $R_p$, and $a$ indicated in each panel
of Figure \ref{fig:probdist_best-fit} and then perform the
$\chi^2$ minimization over $V\sin I_s$ and $\lambda$.

The results are plotted as contour levels in Figure
\ref{fig:probdist_best-fit}.  Here, the two contour curves in each panel
represent the 1 and 2 $\sigma$ levels of the joint probability.
The solid curves plotted along the horizontal and vertical axes are
the probability distributions of $V\sin I_s$ and $\lambda$, respectively,
each of which is the projection of the joint probability distribution
over the other parameter.  Note that in both cases the probability
distribution is well approximated by a Gaussian distribution with
1 $\sigma$ values of $\sigma_{\scriptscriptstyle V\sin I_s}\simeq310$ m
s$^{-1}$ and $\sigma_{\lambda} \simeq 3.\!\!^{\circ}4$.  This result
indicates that the estimated value for $V\sin I_s$ is rather sensitive
to the assumed value of the planetary radius $R_p$, while $\lambda$ can
be estimated reliably even if $R_p$ is not known accurately. This comes
from the fact that the velocity shift is roughly proportional to $V\sin
I_s R_p^2$; however, the projected angle $\lambda$ is sensitive only to
the asymmetry of the radial velocity shift curve. Another important
aspect is that the uncertainty of the prior knowledge of the stellar radius
has little effect on the parameter estimation.

\begin{figure}[htb]
 \epsscale{0.85} \plotone{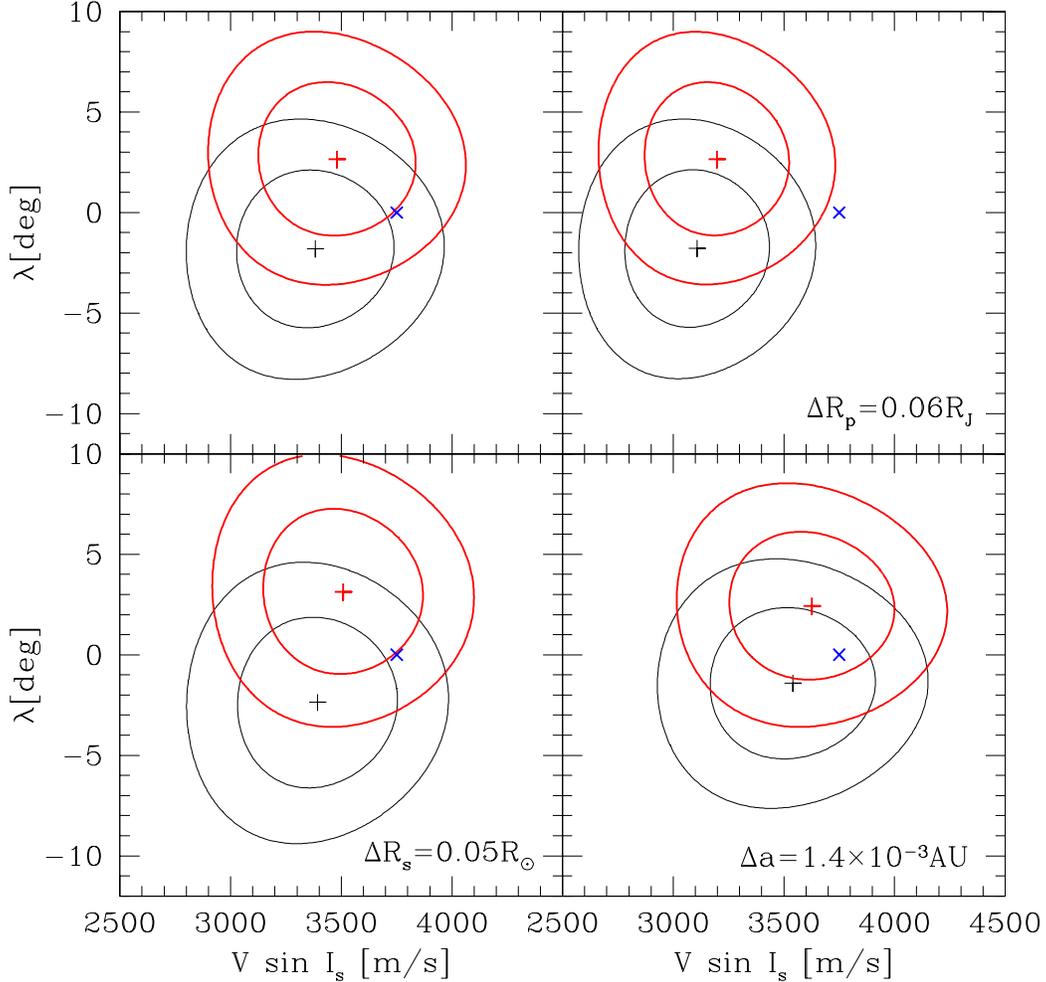} 
\figcaption{Expected constraints on the spin parameters $(V\sin
 I_s,~\lambda)$ from the $\chi^2$ fit to the mock data of radial
 velocity curves for the HD 209458 system. The thin and the thick lines
 show the results obtained from the different mock samples created
 according to the same parameters, as listed in table
 \ref{tab:planetsystem}.  The two contour curves in each line width
 represent the 1 and 2 $\sigma$ confidence levels.  The locations of
 the crosses represent the true values of the spin parameters
 $(V\sin I_s,~\lambda)$, while the locations of the plus signs indicate
 the best-fit parameters, which were estimated under the prior
 assumption for the limb-darkening coefficient and the orbital
 eccentricity as the corect parameters ({\it top left}), $\Delta R_p=0.06\,R_{\rm J}$ ({\it top right}), $\Delta R_s=0.05\,R_\odot$ ({\it
 bottom left}), and $\Delta a=1.4\times 10^{-3}\,{\rm AU}$ ({\it
 bottom right}).  \label{fig:delta_chi^2_contour}}
\end{figure}

\begin{figure}[htbp]
 \epsscale{0.9}
 \plottwo{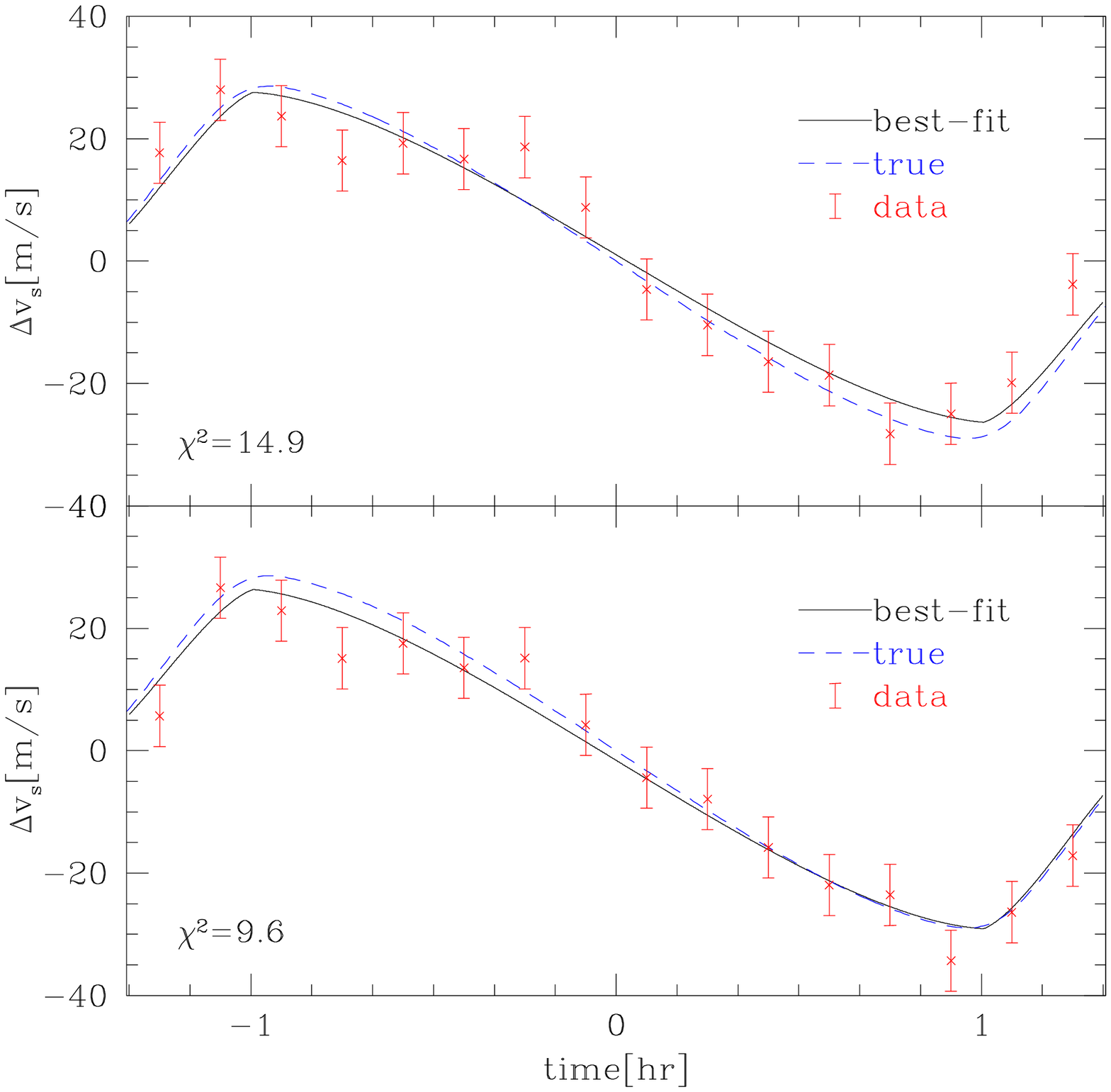}{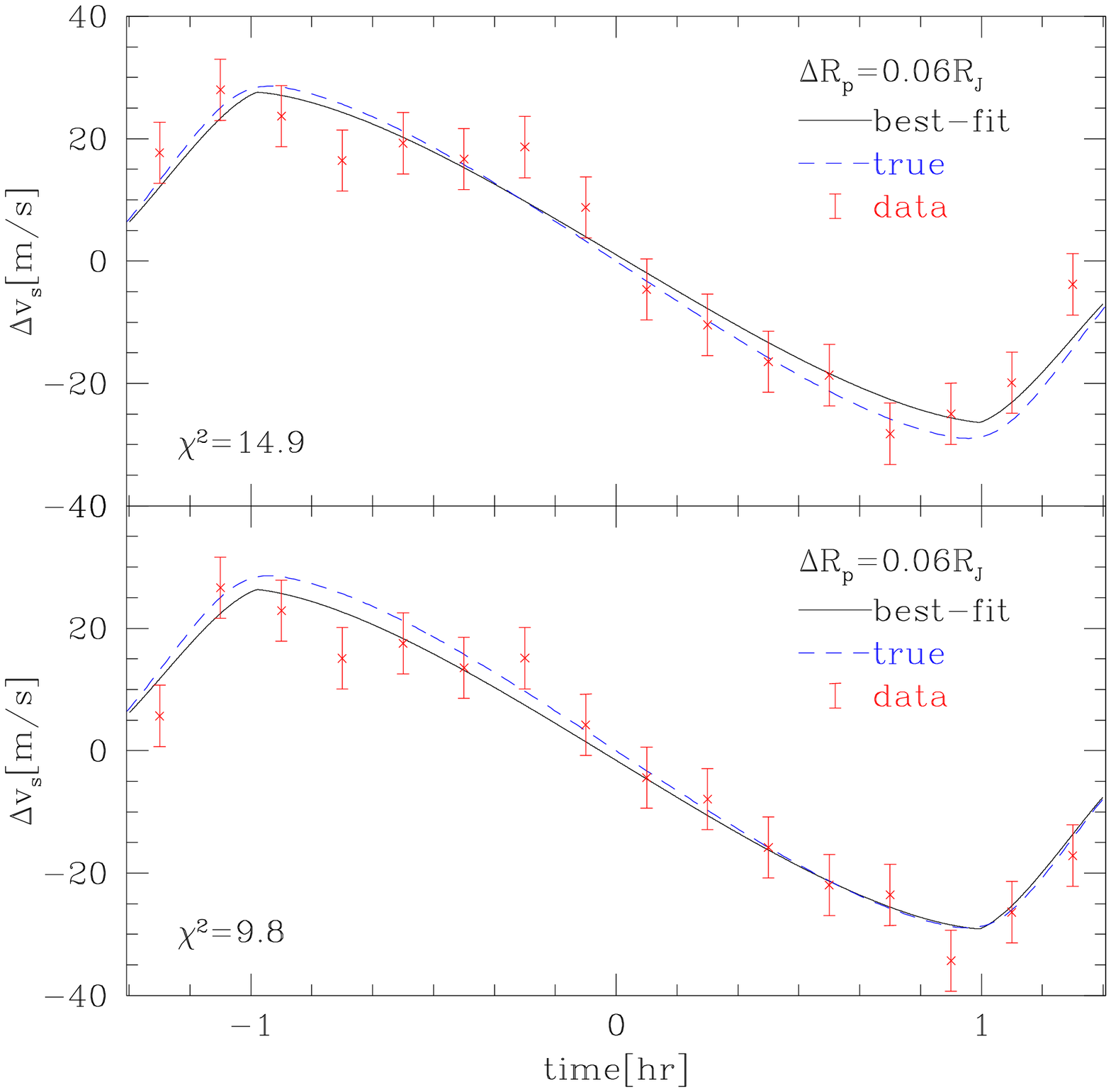}\\
 \begin{minipage}{\textwidth}
  \plottwo{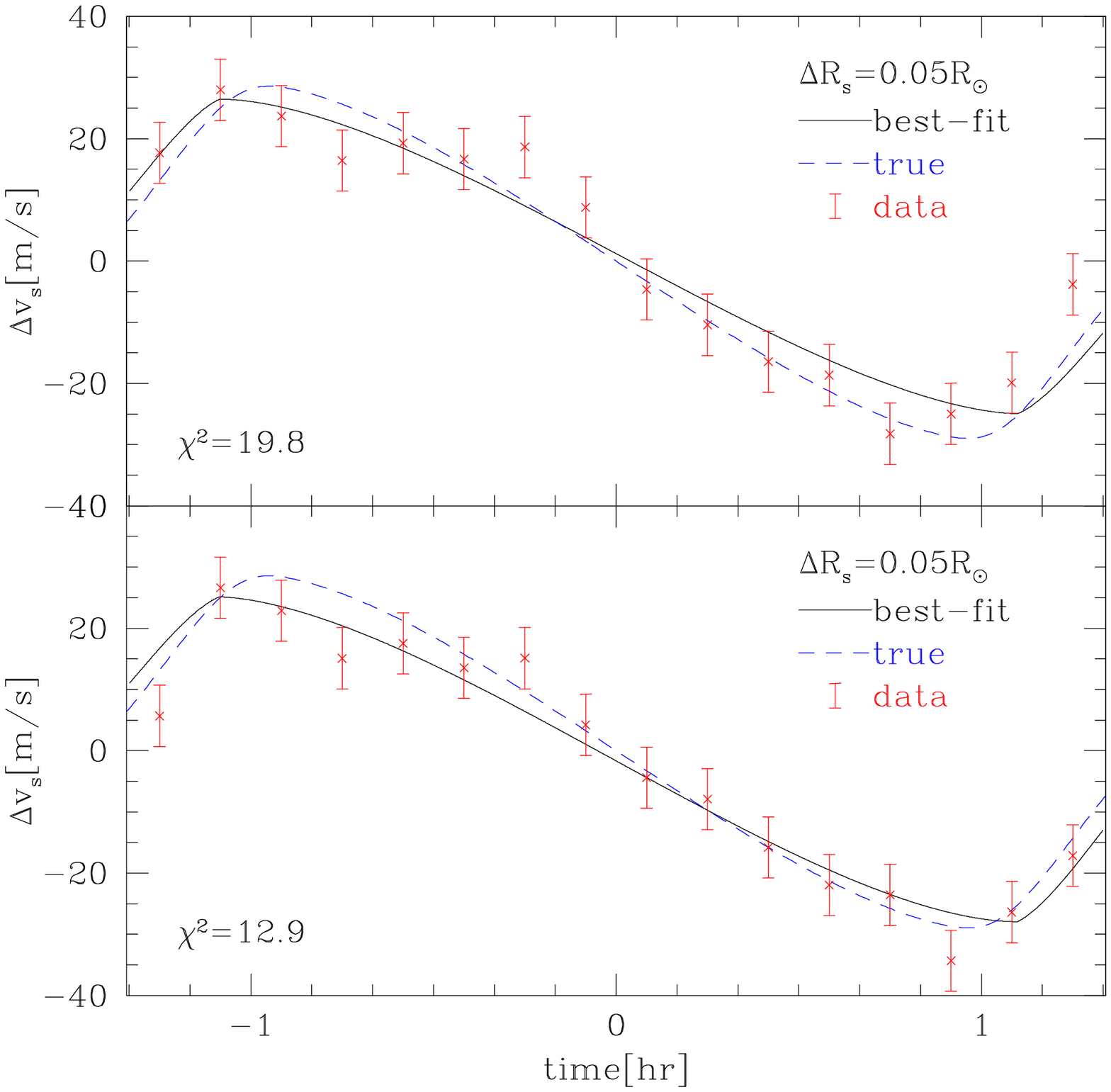}{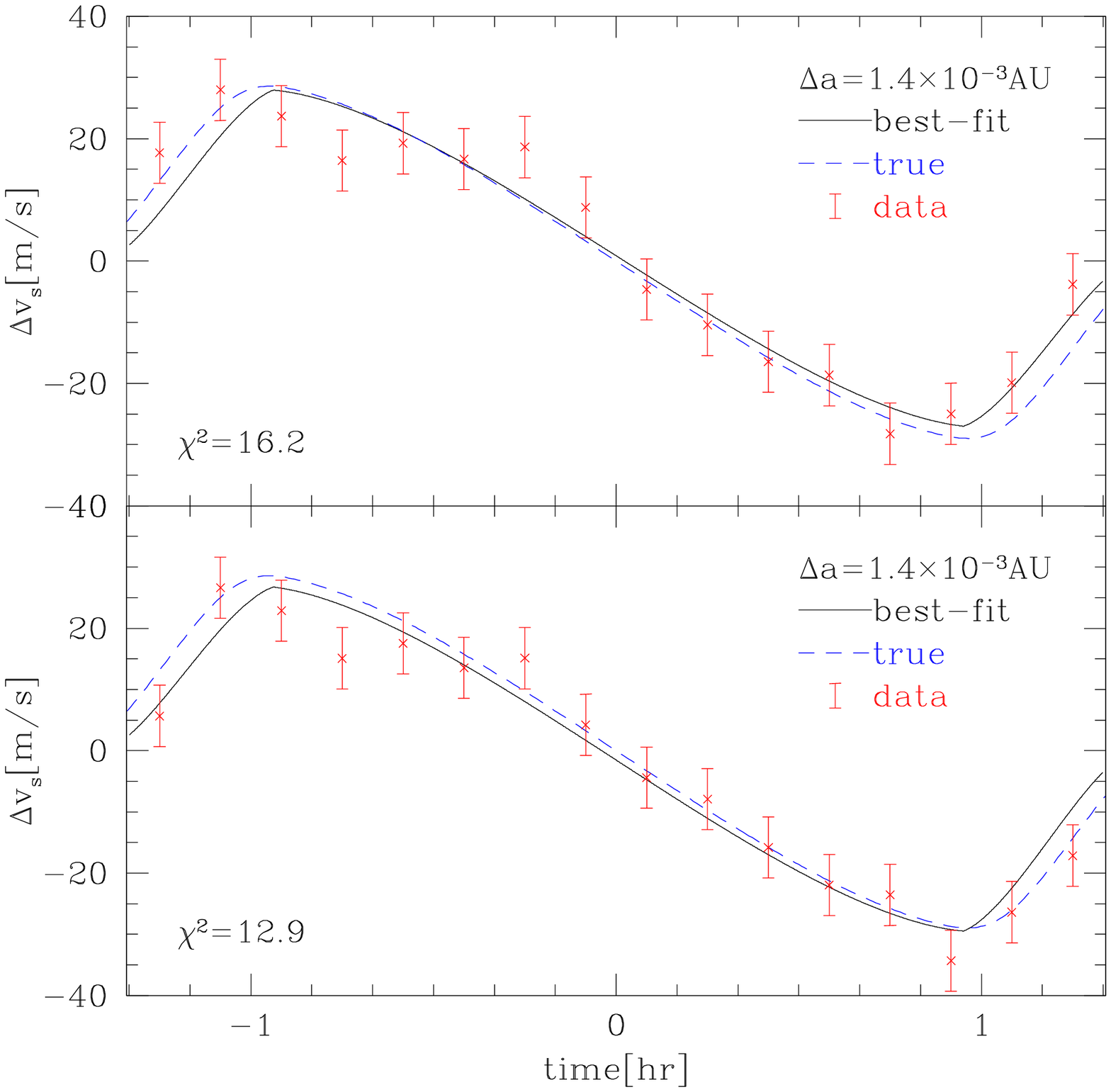}
 \end{minipage} 
 \figcaption{Comparison between the mock data for the radial velocity
 curves used in Fig.\ref{fig:delta_chi^2_contour} and the best-fit
 curves for the parameters $(V\sin I_s,~\lambda)$ ({\it solid curves})
 determined under the prior assumption of $(\epsilon,~e)$ for the fiducial
 parameters ({\it top left}), $\Delta R_p=0.06\,R_{\rm J}$ ({\it
 top right}), $\Delta R_s=0.05=\,R_\odot$ ({\it bottom left}), and $\Delta
 a=1.4\times 10^{-3}\,{\rm AU}$ ({\it lower-right}).  The result depicted
 in the upper and lower window of each panel corresponds to
 the thin and thick contour lines in Fig.\ref{fig:delta_chi^2_contour},
 respectively.
 For comparison, we also plot the correct radial velocity curves in each
 plot ({\it dashed curves}).  \label{fig:demos}}
\end{figure}

The above analysis implies that with a suitably short exposure time, our
formulae provide an unbiased estimation of the spin parameters {\it
statistically}, if we have reasonably accurate prior knowledge of the
other parameters of the system. In reality, however, it may be more
relevant to ask about the reliability of the confidence level of the spin
parameters derived from {\it a single realization}, rather than from the
50,000 realizations. This is related to the above analysis
statistically, but perhaps it is more appropriate to the situation that
one encounters in any observation.  For this purpose, we randomly select
one from the 50,000 realizations and compute 1 and 2 $\sigma$
confidence contours from the {\it relative} confidence levels,
\begin{eqnarray}
&&\Delta\chi^2 \equiv \chi^2(V\sin I_s, \lambda) - 
\chi^2(V\sin I_{s,0}, \lambda_0) ,
\end{eqnarray}
where $V\sin I_{s,0}$ and $\lambda_0$ are the best-fit values.  Figure
\ref{fig:delta_chi^2_contour} shows the estimated parameter regions on
the $V\sin I_s$ versus $\lambda$ plane, and the best-fit values are
indicated by the plus signs in each panel.  The corresponding radial
velocity curves are depicted in Figure \ref{fig:demos}, together with
both the best-fit and the true curves ({\it solid and dashed curves,
respectively}).

Basically, Figure \ref{fig:delta_chi^2_contour} demonstrates that the
spin parameters $V\sin I_s$ and $\lambda$ can be constrained around the
best-fit values with the 1 $\sigma$ errors of $\Delta V\sin
I_s\simeq\pm300$m s$^{-1}$ and $\Delta\lambda\simeq\pm4^{\circ}$, which
greatly improves the uncertainties (by a factor of 4) relative to the
previous result by \citet[see Table \ref{tab:diff_notation}]{Queloz00}.

\section{Conclusions and discussion}
\label{sec:conclusion}

We have discussed a methodology to estimate the stellar spin angular
velocity and its direction angle with respect to the planetary orbit for
transiting extrasolar planetary systems using the RM effect previously
known in eclipsing binary stars \citep{Rossiter24,McLaughlin24,Kopal99}.
In particular we have derived analytic expressions of the radial
velocity anomaly, $\Delta v_s$, which are sufficiently accurate for
extrasolar planetary systems. If the stellar limb darkening is
neglected, the expression is exact. We have extended the result to the
case with limb darkening and obtained approximate but accurate analytic
formulae.  For a typical value of $\gamma=R_p/R_s\sim0.1$, the formulae
reduce to a simple form (eqs.[\ref{eq:dV_s_limbdarkening_A}],
[\ref{eq:approx_W1}], [\ref{eq:approx_W2}],
[\ref{eq:dV_s_limbdarkening_B}], [\ref{eq:approx_W3}], and
[\ref{eq:approx_W4}]):
\begin{eqnarray}
\Delta v_{\rm s}=  \Omega_s X_p \sin I_s \,\,
\frac{ \gamma^2\,\{1-\epsilon(1-W_2)\}}
{ 1-\gamma^2-\epsilon\left\{\frac{1}{3}-\gamma^2\right\}} 
\end{eqnarray}
during the complete transit phase and 
\begin{eqnarray}
\Delta v_{\rm s} &=&  \Omega_s X_p \sin I_s 
\,\,\frac{\displaystyle 
(1-\epsilon)\left\{-z_0\zeta + \gamma^2\cos^{-1}(\zeta/\gamma)\right\} 
+ \frac{\epsilon}{1+\eta_p}\,\, W_4}
{\pi\left(1-\frac{1}{3}\epsilon\right)- 
(1-\epsilon)\left\{\sin^{-1}z_0-(1+\eta_p)z_0
+\gamma^2\cos^{-1}(\zeta/\gamma)\right\}}
\end{eqnarray}
during the egress/ingress phases, where
\begin{eqnarray}
W_2 & = & \frac{(R_s^2-X_p^2-Z_p^2)^{1/2}}{R_s}, \\
W_4 & = & \frac{\pi}{2}\,\,\gamma^{3/2}(2-\gamma)^{1/2}(\gamma-\zeta)\,\,x_c
\,\,\frac{g(x_c\,;\,\eta_p,\,\gamma)}
{g(1-\gamma;\,-\gamma,\,\gamma)},
\end{eqnarray}
where $g(x;a,b)$ is defined in equation (\ref{appen:func_g}).  The
definition and the meaning of the variables in the above expressions are
summarized in Table \ref{tab:notation}.  

The numerical accuracy of the above formulae was checked using a
specific example of the transiting extrasolar planetary system,
HD 209458, and we found that they are accurate within a few percent.  Our
analytic formulae for the radial velocity anomaly are useful in several
ways. One can estimate the planetary parameters much more efficiently
and easily, since one does not have to resort to computationally
demanding numerical modeling. Furthermore, the resulting uncertainties of
the fitted parameters and their correlations are easily evaluated.  To
illustrate these advantages specifically, we performed a parameter
estimation applying the formulae against mock data for HD 209458.
We showed that with precision data obtainable by 8--10 m class
telescopes such as Subaru HDS, our formulae improve the efficiency and
robustness of estimating the spin parameters, $V\sin I_s$ and $\lambda$.
Furthermore, the combined data analysis with asteroseismology
\citep[e.g.,][]{Gizon2003} and/or the correlation between the mean level
 of emission and the rotation period \citep[e.g.,][]{Noyes1985}
may break the degeneracy between $V$ and $I_s$ in extrasolar planetary
systems.

Among the recently discovered transiting extrasolar planetary systems, i.e.,
TrES-1 by the Trans-Atlantic Exoplanet Survey \citep[][]{Alonso} and
OGLE-TR 10, 56, 111, 113, 132 by the Optically Gravitational Lens Event
survey \citep[e.g.,][]{Udalski02a,Udalski02b,Udalski02c,Udalski03,Konacki03,Bouchy,Pont04},
TrES-1 has similar orbital period and mass to those of HD 209458b,
but its radius is smaller.
Thus, it is an interesting target to determine the spin
parameters via the RM effect; if its planetary orbit
and the stellar rotation share the same direction as discovered for the
HD 209458 system, it would be an important confirmation of the current
view of planet formation out of the protoplanetary disk surrounding
the protostar. If not, the result would be more exciting and even
challenge the standard view, depending on the value of the misalignment
angle $\lambda$.

We also note that the future satellites {\it Corot} and {\it Kepler}
will detect numerous transiting planetary systems, most of which will be
important targets for the RM effect in 8 - 10 m
class ground-based telescopes.  We hope that our analytic formulae
presented here will be a useful template in estimating parameters for
those stellar and planetary systems.

Finally, it is interesting to note that the RM effect may even be used as
yet another new detection method of transiting planetary systems.  For
the HD 209458 system, the stellar radial velocity amplitude due to the
Kepler motion is around 100 m s${}^{-1}$. Since the stellar rotation
velocity is around 4 km s${}^{-1}$, the maximum radial velocity anomaly
due to the RM effect is $\sim 40 (\gamma/0.1)^2$m s${}^{-1}$ and thus is
very close to the former. On the other hand, the latter could be
significantly larger if the host star rotates faster, and/or the host
star (the planet) has a smaller (larger) radius.  In extreme cases, the
radial velocity curve, with a periodicity of a few days for instance, is
barely detectable, but the velocity anomaly is quite visible for a few
hours of the transiting phase. Thus, the conventional radial velocity curve
analysis might have missed some of the potentially interesting
spectroscopic signature of transiting planets.  A search for such
specific signatures deserves an attempt against existing or future
spectroscopic samples that do not show any clear conventional feature
of radial velocity periodicity.

In conclusion, we have demonstrated that the radial velocity anomaly due
to the RM effect provides a reliable estimation of spin
parameters. Combining data with the analytic formulae for radial velocity
shift $\Delta v_s$, this methodology becomes a powerful tool in
extracting information on the formation and the evolution of extrasolar
planetary systems, especially the origin of their angular momentum.
Although it is unlikely, we may even speculate that a future RM observation
may discover an extrasolar planetary system in which the stellar spin
and the planetary orbital axes are antiparallel or orthogonal.  This
would have a great impact on the planetary formation scenario, which
would have to invoke an additional effect from possible other planets in the
system during the migration or the capture of a free-floating
planet. While it is premature to discuss such extreme possibilities at
this point, the observational exploration of transiting systems using
the RM effect is one of the most important probes for a better
understanding of the origin of extrasolar planets.

\acknowledgments

We thank Christopher Leigh, Norio Narita, and Edwin Turner for useful
discussions and comments.  We are also grateful to Didier Queloz for
providing the ELODIE data of the radial velocity of HD 209458.  This work
is supported in part by a Grant-in-Aid for Scientific Research from the
Japan Society for Promotion of Science (14102004, 16340053).

\clearpage
\appendix
\section{Approximation of integrals} 
\label{sec:approximation}

In this appendix, the approximate expressions for the integrals 
$W_i~(i=1\sim4)$ defined in section \ref{sec:limb_darkening} are 
derived. Below, we present the results separately for the 
integrals $W_1$ and $W_2$ in Appendix \ref{subsec:W_1_2} 
and for $W_3$ and $W_4$ in Appendix \ref{subsec:W_3_4}.

\subsection{Integrals $W_1$ and $W_2$}
\label{subsec:W_1_2}

First, consider the integral $W_1$. 
For further reduction of the integral, it is convenient to 
rewrite equation (\ref{eq:W_1}) in terms of the new variables
\begin{eqnarray}
(x,z)=R_p\,\sigma(\cos\varphi,\sin\varphi)+(X_p,Z_p), ~~~~~
(X_p,Z_p)=R_s\rho(\cos\theta,\sin\theta).
\label{appen:new_coordinates}
\end{eqnarray}  
Then we obtain 
\begin{eqnarray}
W_1 &=& \frac{1}{\pi}\int_0^1 d\sigma \int_0^{2\pi} d\varphi \,\,
\sigma \sqrt{1-\rho^2\sin^2{(\theta-\varphi)}-
\{\gamma\sigma+\rho\cos{(\theta-\varphi)}\}^2}.
\label{appen:W_1}
\end{eqnarray}
Here, the variable $\rho$ runs from $0$ to $1-\gamma$.
In the above expression, the integral over $\sigma$ is analytically 
expressed, and a part of it is further integrated out. 
The resulting expression becomes 
\begin{eqnarray}
W_1 &=& \frac{2}{3\gamma^2}\,(1-\rho^2)^{1/2}\,
\left(1-\frac{1}{4}\rho^2\right) \,-\, w_A \,+\, w_B, 
\label{appen:W_1_reduced}
\end{eqnarray}
where $w_A$ and $w_B$ are given by 
\begin{eqnarray}
w_A &=& \frac{1}{\pi\,\gamma^2}\,\,\left[
\frac{1}{3}\int_0^{2\pi} d\varphi (1-\rho^2-\gamma^2-2\rho\gamma\
\cos{\varphi})^{3/2}
\right. \cr
&&\left.~~~~~~~~~~~~~ +\frac{1}{2}
\int_0^{2\pi} d\varphi\,\, (\rho^2\cos\varphi+\rho\gamma)\cos\varphi
(1-\rho^2-\gamma^2-2\rho\gamma\cos\varphi)^{1/2}\right],
\label{appen:w_A}
\\
w_B &=&\frac{1}{2\pi\gamma^2}\int_0^{2\pi}d\varphi\,\,\rho\cos\varphi 
(1-\rho^2\sin^2\varphi)
\cr
&&~~~~~~~~~~~~~~~~~~\times\,\left\{
\sin^{-1}\left(\frac{\rho\cos\varphi}{\sqrt{1-\rho^2\sin^2\varphi}}\right)
-\sin^{-1}\left(\frac{\gamma+\rho\cos\varphi}
{\sqrt{1-\rho^2\sin^2\varphi}}\right)\right\}.
\label{appen:w_B}
\end{eqnarray}
Note that the integrals $w_A$ and $w_B$ are analytically expressed in
terms of the complete elliptic integrals of
the first kind $K(m)$, the second kind $E(m)$, and the third kind
$\Pi(n,m)$:
\begin{eqnarray}
w_A &=& \frac{1}{45\pi\gamma^4}\{1-(\gamma+\rho)^2\}^{1/2}
\cr
&& \times \,\,\,\left[
\left\{3(1-\rho^2)^2-\gamma^2(71-17\rho^2)+68\gamma^4\right\}\,E(m)
\right.
\cr
&&~~~~~~~~~~~
\left.-\{(\gamma-\rho)^2-1\}(3\rho^2-3+8\gamma^2)\,K(m)
\right],
\label{appen:exact_w_A}
\end{eqnarray}
\begin{eqnarray}
 w_B &=& \frac{2}{3\gamma^2}\left[1-(1-\rho^2)^{1/2}\left(1-\frac{\rho^2}{4}\right)\right]\nonumber\\
 && +\frac{1}{45\pi\gamma^4[1-(\gamma+\rho)^2]^{1/2}}\nonumber\\
 && ~~\times\left\{\left[2\gamma^6+(29-7\rho^2)\gamma^4+(26-34\rho^2+8\rho^4)\gamma^2+3(1-\rho^2)^3\right]K(m)\right.\nonumber\\
 && ~~~~~~\left.+\left[1-(\rho+\gamma)^2\right]\left[2\gamma^4+(31-7\rho^2)\gamma^2-3(1-\rho^2)^2\right]E(m)\right\}\nonumber\\
 && -\frac{2}{3\pi\gamma^2[1-(\gamma+\rho)^2]^{1/2}}\left\{\left[1-n(\rho+\gamma)/2\rho\right]\Pi(n,m)+\rm{c.c.}\right\},
\label{appen:exact_w_B}
\end{eqnarray}
where c.c. denotes the complex conjugate,
$n=2\rho\left[\rho-i(1-\rho^2)^{1/2}\right]$, and
$m=-4\gamma\rho/[1-(\gamma+\rho)^2]$.  To evaluate equations
(\ref{appen:exact_w_A}) and (\ref{appen:exact_w_B}), careful treatments
are required at the edge $\rho=1-\gamma$, where the argument of the
elliptic integral and coefficients of $K(m)$ and $\Pi(n,m)$ in $w_B$
apparently diverge.  Since we are concerned with planetary systems
with a small ratio $R_p\ll R_s$, it is practically useful to derive
the approximate expressions.  In this case, we do not have to use the
complicated expressions (\ref{appen:exact_w_A}) and
(\ref{appen:exact_w_B}), but can expand the integrands
(\ref{appen:W_1}) and/or (\ref{appen:w_A}), (\ref{appen:w_B}) in powers
of $\gamma=R_p/R_s$.  Then
each term in the expansion with respect to $\gamma$ can be analytically
integrated.  The results up to the second order in $\gamma$ become
\begin{eqnarray}
w_A &\simeq& 
\frac{2}{3\gamma^2}\,\,(1-\rho^2)^{1/2}\left(1-\frac{1}{4}\rho^2\right)
\,\,-\,\,\frac{1}{16}\,\,\frac{15\rho^4-28\rho^2+16}{(1-\rho^2)^{3/2}}
\cr
&&~~~~~~~~~~~~~~~~~~~-\,\gamma^2\,\,\frac{1}{128}\,\,
\frac{17\rho^6-64\rho^4+104\rho^2-32}{(1-\rho^2)^{7/2}}
\,\,+\,\,{\cal O}(\gamma^4), \\
w_B &\simeq& 
-\frac{1}{4}\,\,\frac{\rho^2(1-\rho^2/4)}{(1-\rho^2)^{3/2}}\,\,-\,\,
\gamma^2\,\,\frac{1}{128}\,\,\frac{\rho^2(24+\rho^4)}{(1-\rho^2)^{7/2}}
\,\,+\,\,{\cal O}(\gamma^4).
\end{eqnarray}
Substituting these expressions into equation (\ref{appen:W_1_reduced}) and 
collecting the terms in powers of $\gamma$,  
we finally obtain 
\begin{equation}
W_1 \simeq 
(1-\rho^2)^{1/2}\,\,-\,\,\gamma^2\,\frac{2-\rho^2}{8(1-\rho^2)^{3/2}} +
{\cal O}(\gamma^4).
\label{appen:approx_W_1}
\end{equation}

Next, consider the integral $W_2$, whose analytic expression is also 
obtained through the same procedure as mentioned above. 
Using equations (\ref{appen:new_coordinates}), 
one writes equation (\ref{eq:W_2}) as 
\begin{equation}
W_2 = W_1 + w_C,
\label{eq:W_2_reduced}
\end{equation}
where $w_C$ is given by 
\begin{equation}
w_C = \frac{\gamma}{\pi\,\rho\cos\theta}\,\,
\int_0^1 d\sigma \int_0^{2\pi} d\varphi \,\,
\sigma^2 \cos\varphi \sqrt{1-\rho^2\sin^2{(\theta-\varphi)}-
\{\gamma\sigma+\rho\cos{(\theta-\varphi)}\}^2}. 
\end{equation}
In the above expression, the integral over $\sigma$ is analytically 
performed, and the resulting expression for the integrand is expanded in 
powers of $\gamma$. Further integrating it over $\varphi$, we obtain 
\begin{equation}
w_C \simeq -\gamma^2\,\,\frac{1}{4(1-\rho^2)^{1/2}} + {\cal O}(\gamma^4).
\end{equation}
Thus, the perturbative expansion for $W_2$ becomes 
\begin{equation}
W_2 \simeq 
(1-\rho^2)^{1/2}\,\,-\,\,\gamma^2\,\frac{4-3\rho^2}{8(1-\rho^2)^{3/2}} +
{\cal O}(\gamma^4).
\label{appen:approx_W_2}
\end{equation}

In Figure \ref{fig:integral_W1_W2}, to check the validity of the 
perturbation results, the approximate expressions for the integrals 
$W_1$ and $W_2$   
are plotted as a function of $\rho$. The results are then compared with 
those obtained from the numerical integration. As is expected, 
the perturbative expressions (\ref{appen:approx_W_1}) and 
(\ref{appen:approx_W_2}) give a quite accurate approximation as long as 
the ratio of the planetary radius to the stellar radius is small. Note that  
the approximation is even better for the slightly larger value $\gamma=0.3$, 
with a few percent level for the fractional error.

\subsection{Integrals $W_3$ and $W_4$}
\label{subsec:W_3_4}

As for the integrals $W_3$ and $W_4$ given by equations (\ref{eq:W_3}) and 
(\ref{eq:W_4}), one can partially evaluate the integrals 
with the knowledge of the integral region (\ref{eq:range_of_int}).   
The resulting forms are summarized as
\begin{eqnarray}
W_3 &=& \frac{\pi}{6}(1-x_0)^2(2+x_0) + 
\int_{x_0+\zeta-\gamma}^{x_0} d\tilde{x}\,\, g(\tilde{x};\,\eta_p,\,\gamma),
\label{appen:reduced_W_3}
\\
W_4 &=& \frac{\pi}{8}(1-x_0^2)^2 + 
\int_{x_0+\zeta-\gamma}^{x_0} d\tilde{x}\,\,\tilde{x} \,\,
g(\tilde{x};\,\eta_p,\,\gamma),
\label{appen:reduced_W_4}
\end{eqnarray}
where we defined the function $g(\tilde{x};\,\eta_p,\,\gamma)$ as
\begin{eqnarray}
g(\tilde{x};\,\eta_p,\,\gamma) &\equiv& (1-\tilde{x}^2)\sin^{-1}
\left\{\frac{\gamma^2-(\tilde{x}-1-\eta_p)^2}{1-\tilde{x}^2}\right\}^{1/2}
\cr
&& ~ + \sqrt{2(1+\eta_p)(x_0-x)\{\gamma^2-(\tilde{x}-1-\eta_p)^2\}}.
\label{appen:func_g}
\end{eqnarray}
Since the above one-dimensional integrals cannot be evaluated
analytically, one may derive an approximate expression applicable 
to the $\gamma\ll1$ cases. 
Note, however, that a naive treatment by the perturbative expansion 
regarding $\gamma$ as a small parameter can break down 
at $\eta_p\simeq-\gamma$. Even in the $\eta_p>\gamma$ case, 
perturbative expression gives a worse approximation. 
For an accurate evaluation of the 
integrals, a more dedicated treatment other than the perturbative expansion 
is required. One clever approach, which we adopt here,
is to replace the function $g$ with
\begin{equation}
g(\tilde{x};\,\eta_p,\,\gamma)~~ \longrightarrow ~~ 
g(x_c;\,\eta_p,\,\gamma) \,\,
\sqrt{\frac{(a-\tilde{x})(\tilde{x}-b)}{(a-x_c)(x_c-b)}},
\end{equation}
where we set $a=x_0+\zeta-\gamma$, $b=x_0$, and $x_c=(a+b)/2$,
and to use the integral formula
\begin{equation}
\int_a^b dx \sqrt{\frac{(a-x)(x-b)}{(a-x_c)(x_c-b)}} =\frac{\pi}{4}(b-a).
\end{equation}
Then, one can approximate the one-dimensional integrals in equations 
(\ref{appen:reduced_W_3}) and (\ref{appen:reduced_W_4}) as
\begin{eqnarray}
\int_{x_0+\zeta-\gamma}^{x_0} d\tilde{x}\,\,
g(\tilde{x};\,\eta_p,\,\gamma) &~~\longrightarrow~~&
\frac{\pi}{4}\,\,(\gamma-\zeta)\,\,g(x_c;\,\eta_p,\,\gamma),
\label{appen:formula_1}
\\
\int_{x_0+\zeta-\gamma}^{x_0} d\tilde{x}\,\,\tilde{x} \,\,
g(\tilde{x};\,\eta_p,\,\gamma) &~~\longrightarrow~~&
\frac{\pi}{8}\,\,(\gamma-\zeta)(2x_0+\zeta-\gamma)\,\,
g(x_c;\,\eta_p,\,\gamma). 
\label{appen:formula_2}
\end{eqnarray}
As long as the planetary transit system has $\gamma\ll 1$, 
the above expression in fact gives an accurate prescription. Note,  
however, that a naive use of the formulae 
(\ref{appen:formula_1}) and (\ref{appen:formula_2}) leads to 
inconsistent radial velocity curves, 
which do not satisfy the junction condition at $\eta_p=-\gamma$: 
\begin{equation}
W_3 = \pi\, \gamma^2\,\, W_1|_{\rho=1-\gamma}, ~~~~~
W_4 = \pi\, \gamma^2(1-\gamma)\,\, W_2|_{\rho=1-\gamma}. 
\label{appen:junction}
\end{equation}

To preserve the consistency,  
we modify  equations (\ref{appen:formula_1}) 
and (\ref{appen:formula_2}) so as to satisfy the junction condition 
(\ref{appen:junction}) by multiplying the numerical factor. 
The final expressions for $W_3$ and $W_4$ become
\begin{eqnarray}
W_3&\simeq& \frac{\pi}{6}(1-x_0)^2(2+x_0) + 
\frac{\pi}{2}\,\,\gamma(\gamma-\zeta)\,\,
\frac{g(x_c;\,\eta_p,\,\gamma)}
{g(1-\gamma;\,-\gamma,\,\gamma)}\,\,
W_1(1-\gamma),
\label{appen:approx_W_3}
\\
W_4&\simeq& \frac{\pi}{8}(1-x_0)^2(1+x_0)^2 + 
\frac{\pi}{2}\,\,\gamma(\gamma-\zeta)\,\,x_c\,\,
\frac{g(x_c;\,\eta_p,\,\gamma)}
{g(1-\gamma;\,-\gamma,\,\gamma)}\,\,W_2(1-\gamma).
\label{appen:approx_W_4}
\end{eqnarray}

\begin{figure}[t]
\epsscale{0.9}
 \plottwo{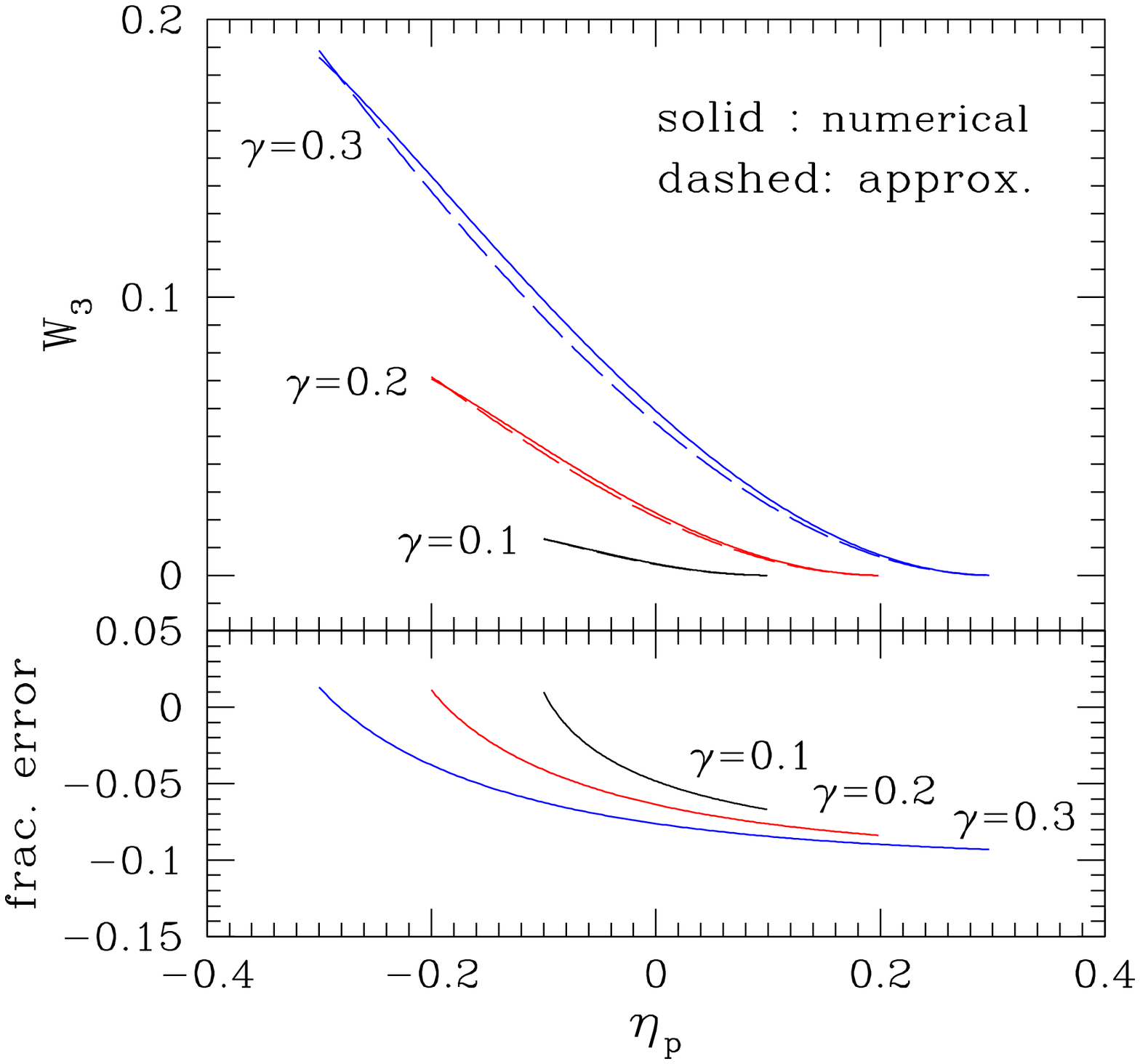}{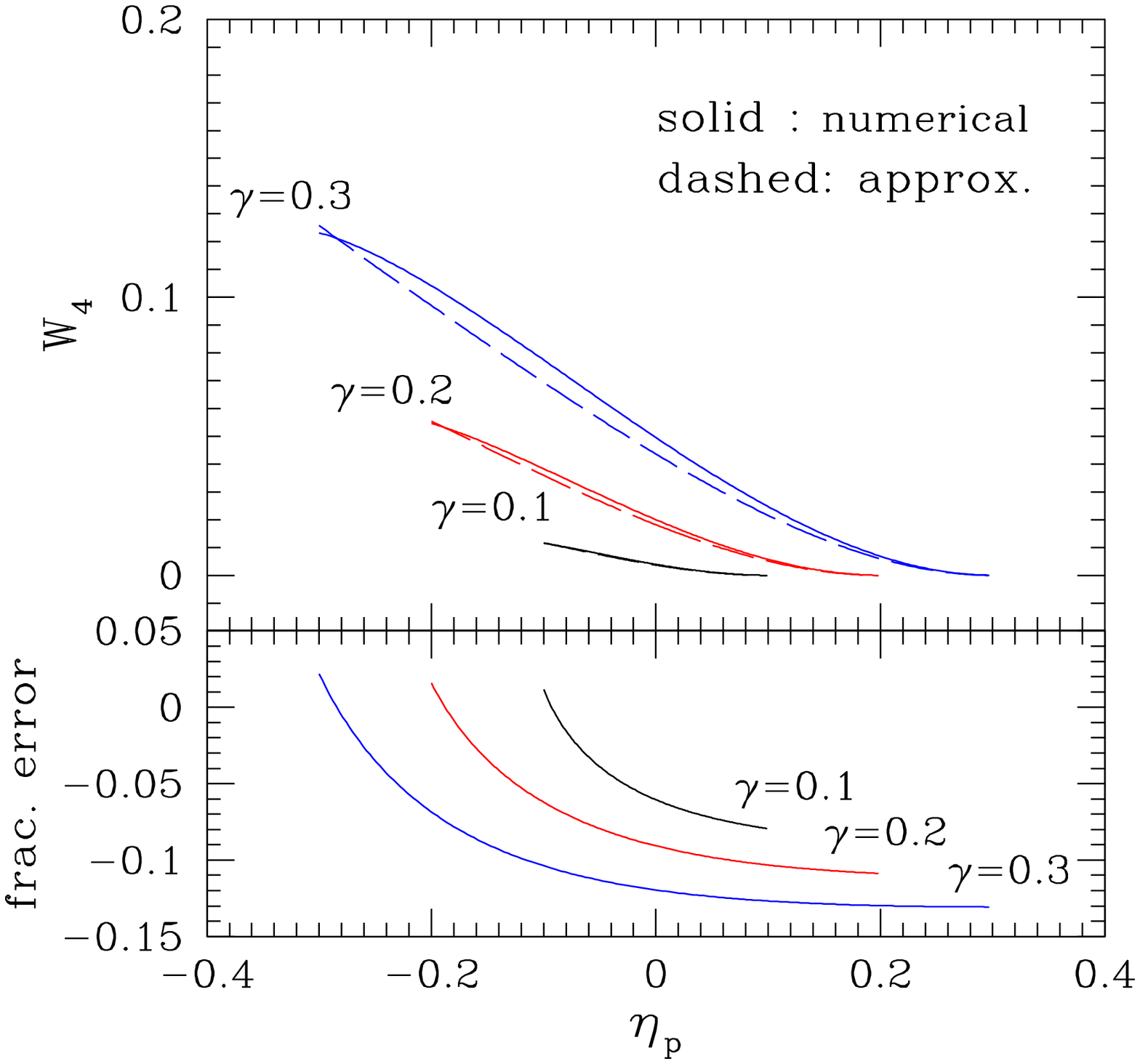}
 \figcaption{
 {\it Top windows}: Approximate vs. numerical evaluations of the 
 integrals $W_3$ ({\it left}) and $W_4$ ({\it right}) 
 as a function of $\eta_p$. The solid lines 
 represent the numerical evaluation of equations 
 (\ref{appen:reduced_W_3}) and (\ref{appen:reduced_W_4}), 
 while the dashed lines are the approximation based on the perturbative 
 expansions (\ref{appen:approx_W_3}) and 
 (\ref{appen:approx_W_4}). Note that the variable $\eta_p$ 
 runs from $-\gamma$ to $+\gamma$. {\it Bottom windows}: fractional errors 
 $W^{\rm (approx)}/W^{\rm (num)}-1$ for the integrals $W_3$ ({\it left}) and  
 $W_4$ ({\it right}). 
 \label{fig:integral_W3_W4}} 
\end{figure}

In figure \ref{fig:integral_W3_W4}, substituting equations 
(\ref{appen:approx_W_1}) and (\ref{appen:approx_W_2}) into  
the above results, the approximate expressions for $W_3$ and $W_4$ are 
depicted as a function of $\eta_p$. Although it is a tricky treatment, 
it turns out that the expressions (\ref{appen:approx_W_3}) and 
(\ref{appen:approx_W_4}) give an accurate approximation  
for the range of our interest. Compared to the integrals $W_1$ and $W_2$, 
the fractional error of the approximations seem slightly large;
however, the contribution of the integrals $W_3$ and $W_4$ 
to the radial velocity shift $\Delta v_s$ is relatively 
small. With a typical parameter $\gamma\sim 0.1$, 
the resulting fractional error remains only a few percent.

\clearpage


\begin{deluxetable}{ccl}
 \tablecaption{List of notation
\label{tab:notation}}
 \tablehead{\colhead{Variables} & \colhead{Definition} &
 \colhead{Meaning} }
 \startdata
 \cutinhead{Orbital Parameters}
 $m_p$ & Sec.\ref{sec:radvel_non-transit} & Planet mass \\
 $m_s$ & Sec.\ref{sec:radvel_non-transit} & Stellar mass \\
 $ a $ & Fig.\ref{fig:orbit-topview} & Semimajor axis \\
 $ e $ & Fig.\ref{fig:orbit-topview} & Eccentricity of planetary orbit \\
 $\varpi $ & Fig.\ref{fig:orbit-topview} & Negative longitude 
	of the line of sight \\
 $i$ & Fig.\ref{fig:spin-los} & Inclination between normal direction of 
orbital plane and $y$-axis \\
 $r_p$ & Eq.[\ref{eq:rp}] & Distance between star and planet 
	(see Fig.\ref{fig:orbit-topview}) \\
 $f$ & Eq.[\ref{eq:true_anomaly}] & True anomaly	
	(see Fig.\ref{fig:orbit-topview}) \\
 $E$ & Eq.[\ref{eq:true_anomaly}] & Eccentric anomaly \\
 $n$ & Eq.[\ref{eq:mean_motion}] & Mean motion \\
 $M$ & Eq.[\ref{eq:mean_anomaly}] & Mean anomaly \\
\cutinhead{Internal Parameters of Star and Planet}
 $I_s$ & Fig.\ref{fig:spin-los} & Inclination between stellar spin axis 
	and $y$-axis \\
 $\lambda$ & Fig.\ref{fig:orbit-xz} & Angle between $z$-axis and 
	normal vector $\hat{\bm{n}}_p$ on $(x,z)$-plane \\
{\bf $\Omega_s$} & Eq.[\ref{eq:angular_veloc}] & Angular velocity of star 
	(see Fig.\ref{fig:spin-los}) \\
$R_s$ & Sec.\ref{sec:without_darkening} & Stellar radius \\
$R_p$ & Sec.\ref{sec:without_darkening} & Planet radius \\
$\epsilon$ & Eq.[\ref{eq:I_limb_darkening}] & Limb darkening parameter \\
$V$ & Sec.\ref{sec:HD209458} & Stellar surface velocity, $R_s\Omega_s$ \\
\cutinhead{Mathematical Notation}
$\bm{X}_p$ & Sec.\ref{sec:without_darkening} & Position of the planet \\
$\gamma$ & Eq.[\ref{eq:v_s_complete}] &
Ratio of planet radius to stellar radius, $R_p/R_s$ \\
$\eta_p$ & Eq.[\ref{eq:def_eta_p}] & See Fig.\ref{fig:ingress} \\
$x_0$ & Eq.[\ref{eq:x0z0}] & See Fig.\ref{fig:ingress} \\
$z_0$ & Eq.[\ref{eq:x0z0}] & See Fig.\ref{fig:ingress} \\
$\zeta$ & Eq.[\ref{eq:zeta}] & See Fig.\ref{fig:ingress} 
 \enddata
 \\
\end{deluxetable}
\begin{deluxetable}{ccc}
 \tablecaption{Parameters of the HD 209458 system
 \label{tab:planetsystem}}
 \tablehead{\colhead{Parameters} & \colhead{Estimated Values} &
 \colhead{Fractional Errors (\%)}}
 \startdata
 $m_s$ & $1.1\pm0.1$~~M$_\odot$\tablenotemark{a}&9.1\\
 $m_p$ &$0.69\pm0.05$~~M$_J$\tablenotemark{a} &7.2\\
 $e$ & 0,\tablenotemark{b}~- 0.1\tablenotemark{c} & \nodata\\
 $a$ & $0.0468$\tablenotemark{a}~$\pm0.0014$\tablenotemark{d}~~AU & 3.0\\
 $\varpi$ &$100^{\circ~}$\tablenotemark{c} &\nodata\\
 $P_{\rm orb}$ &$3.52474\pm0.00007$ ~days\tablenotemark{a} &0.002\\
 $R_s$ & $1.146\pm0.050$ ~~R$_\odot$\tablenotemark{a} &4.4\\
 $R_p$& $1.347\pm0.060$ ~~R$_J$\tablenotemark{a} &4.5\\
 $i$ & $86^\circ.1\pm0^\circ.1$\tablenotemark{b}&0.1\\
 $\epsilon$ & $0.64\pm0.03$\tablenotemark{a}&4.7\\
 \enddata
 \tablenotetext{a}{From \citet{Brown01}.}
 \tablenotetext{b}{From http://www.obspm.fr/encycl/HD209458.html.}
 \tablenotetext{c}{From http://exoplanets.org/esp/hd209458/hd209458.html.}
 \tablenotetext{d}{This error is calculated from those of $m_s$, $m_p$,
 and $P_{orb}$.}
\end{deluxetable}
\begin{deluxetable}{ccc}
 \tablecaption{Spin Parameters Derived by \citet{Queloz00} and
 Notational Differences
 \label{tab:diff_notation}}
 \tablehead{\colhead{This Paper}&\colhead{\citet{Queloz00}}& \colhead{Best-Fit Solutions\tablenotemark{a}}}
 \startdata
 $V\sin I_s$ &$v\sin i$ &$3.75 \pm 1.25$~\tablenotemark{b}\\
 $\mbox{sgn}(\lambda)\,\cos^{-1}\{\cos\lambda \sin i\}$& $\alpha$&
 $\pm 3.9 {+18 \atop -21}$ \tablenotemark{c}\\
 $\lambda$& 
 $\mbox{sgn}(\alpha)\cos^{-1}\left[\frac{\cos\alpha}{\left(1-\cos^2\Omega_p\sin^2\alpha\right)^{1/2}}\right]$ &
 0 for $(\alpha,\Omega_p)=(3.9, 0)$\tablenotemark{c}\\
 && 21.7 for $(\alpha,\Omega_p)=(22, 100)$\tablenotemark{c}\\
 && $-24.7$ for $(\alpha,\Omega_p)=(-25, 80)$\tablenotemark{c}\\
 \enddata
 \tablenotetext{a}{The parameter $\Omega_p$ is related to the
 inclination angle and is constrained through\\
 $\cos\Omega_p=-\cos(86.\!\!^{\circ}1)/\cos\alpha$.}
 \tablenotetext{b}{In units of km s$^{-1}$.}
 \tablenotetext{c}{Units for $\alpha$, $\lambda$, $\Omega_p$ are degrees.}
\end{deluxetable}

\clearpage


\begin{thebibliography}{}

\bibitem[Alonso et al.(2004)]{Alonso}
Alonso, R., Bron, T.M., Torres, G., Latham, D.W. Sozzetti, Al., 
Mandushev, G., Belmonte, J.A., Charbonneau, D., Deeg, H.J., 
Dunham, E.W., O'Donovan, F.T., and Stefanik, R.P. 
2004, \apj, 613, L153

\bibitem[Bouchy et al.(2004)]{Bouchy}
Bouchy, F., Pont, F., Santos, N.C.,  Melo, C., Mayor, M., 
Queloz, D., and  Udry, S. 2004, \aap, 421, L13 

\bibitem[Brown et al.(2001)]{Brown01} Brown, T.M. et al., 2001, \apj,
552, 699.

\bibitem[Charbonneau et al.(1998)]{Charbonneau98} Charbonneau, D.,
Jha, S., and Noyes, R. W., 1998, \apj 507, L153.

\bibitem[Charbonneau et al.(1999)]{Charbonneau99} Charbonneau, D. et
al., 1999, \apj 522, L145.

\bibitem[Charbonneau et al.(2000)]{Charbonneau00} Charbonneau, D. et
al., 2000, \apj 529, L45.

\bibitem[Claret (2000)]{Claret00} Claret, A. 2000, \aap, 363, 1081

\bibitem[Gizon \& Solanki (2003)]{Gizon2003} Gizon, L., \&  Solanki, S, K. 2003, \apj, 
589, 1009.

\bibitem[Henry et al.(2000)]{henry00} Henry, G.W. et al., 2000, \apj,
529, L41.

\bibitem[Hosokawa (1953)]{Hosokawa53} Hosokawa, Y. 1953, \pasj, 5, 88.

\bibitem[Konacki et al.(2003)]{Konacki03} Konacki, M., Torres, G., Jha,
S. and Sasselov, D. D., 2003, \nat, 421, 507.

\bibitem[Kopal (1942)]{Kopal42} Kopal, Z.
 1942, Proc. U.S. Natl. Acad. Sci., 28, 133

\bibitem[Kopal (1945)]{Kopal45} Kopal, Z.
 1945, Proc. Am. Phil. Soc. 89, 517

\bibitem[Kopal (1990)]{Kopal99} Kopal, Z. 1990, Mathematical 
Theory of Stellar Eclipses (Dordrecht: Kluwer).

\bibitem[Lin et al.(1996)]{Lin96} Lin, D.N.C., Bodenheimer, P.,
Richardson, D.C. 1996, \nat, 380, 606

\bibitem[McLaughlin (1924)]{McLaughlin24} 
McLaughlin, D. B. 1924, \apj, 60, 22

\bibitem[Murray \& Dermott(1999)]{SSD} Murray, C.D. and Dermott, S.F., 1999,
 Solar System Dynamics (Cambridge: University of Cambridge Press)

\bibitem[Noyes et al.(1985)]{Noyes1985} Noyes, R.W., Hartmann, L.W., 
Baliunas, S.L., Duncan, D.K., and Vaughan, A.H. 1985, \apj, 279, 763.

\bibitem[Petrie (1938)]{Petrie38} Petrie, R. M. 1938, Publ. Dominion
			     Astrophys. Obs., 7, 133

\bibitem[Pollack et al.(1996)]{Pollack96} Pollack, J.B., Hubickyj, O.,
Bodenheimer, P., et al. 1996, Icarus, 124, 62 

\bibitem[Pont et al.(2004)]{Pont04} Pont, F., Bouchy, F., 
Queloz, D., Santos, N.~C., Melo, C., Mayor, M., \& Udry, S.\ 2004, \aap, 
426, L15

\bibitem[Queloz et al.(2000)]{Queloz00} Queloz, D., Eggenberger, A.,
Mayor, M., Perrier, C., Beuzit, J.L., Naef, D., Sivan, J.P., and Udry,
S., 2000, \aap, 359, L13.

\bibitem[Rossiter (1924)]{Rossiter24} Rossiter, R. A. 1924, \apj 69, 15. 

\bibitem[Santos et al.(2004)]{HARPS} Santos, N.C., Bouchy, F., Mayor, M.,
et al. 2004, \aap, in press (astro-ph/0408471)

\bibitem[Schlesinger(1909)]{Schlesinger1909} Schlesinger, F. 1910
Publ. Allegheny Obs., 1, 123

\bibitem[Snellen(2004)]{Snellen04} Snellen, I. A. G. 2004, 
\mnras, 353, L1

\bibitem[Udalski et al.(2002a)]{Udalski02a} Udalski, A., Paczy\'nski, B.,
\.{Z}ebr\'{u}n, K., Szyma\'{n}ski, M., Kubiak, M., Soszy\'{n}ski, I.,
Szewczyk, O., Wyrzykowski, \L., and Pietrzy\'{n}ski, G., 2002a, \actaa 52, 1

\bibitem[Udalski et al.(2002b)]{Udalski02b} Udalski, A.,
\.{Z}ebr\'{u}n, K., Szyma\'{n}ski, M., Kubiak, M., Soszy\'{n}ski, I.,
Szewczyk, O., Wyrzykowski, \L., and Pietrzy\'{n}ski, G., 2002b, \actaa 52, 115

\bibitem[Udalski et al.(2002c)]{Udalski02c} Udalski, A., Szewczyk, O.,
\.{Z}ebr\'{u}n, K., Pietrzy\'{n}ski, G., Szyma\'{n}ski, M., Kubiak, M.,
Soszy\'{n}ski, I., and Wyrzykowski, \L., 2002c, \actaa 52, 317

\bibitem[Udalski et al.(2003)]{Udalski03} Udalski, A., Pietrzy\'{n}ski, G.,
Szyma\'{n}ski, M., Kubiak, M., \.{Z}ebr\'{u}n, K., Soszy\'{n}ski, I.,
Szewczyk, O., and Wyrzykowski, \L., 2003, \actaa 53, 133

\bibitem[Winn et al.(2004)]{Winn2004}
Winn, J.N., Suto, Y., Turner, E.L., Narita, N., Frye, B.L., 
Aoki, W., Sato, B., \& Yamada, T. 2004, \pasj, 56, 655

\end{thebibliography}
\end{document}